\documentclass[apj,twocolappendix]{emulateapj}
\usepackage{natbib,gensymb,wasysym}
\shorttitle{Acceleration and escape of matter in the oblique black hole magnetosphere}
\shortauthors{Kop\'{a}\v{c}ek \& Karas}

\begin{document}
\title{Near-horizon structure of escape zones of electrically charged particles around weakly magnetized rotating black hole. II. Acceleration and escape in the oblique magnetosphere}
\email{kopacek@ig.cas.cz}
\author{Ond\v{r}ej Kop\'{a}\v{c}ek}
\affiliation{Astronomical Institute, Czech Academy of Sciences, Bo\v{c}n\'{i} II, CZ-141\,00~Prague, Czech~Republic}
\affiliation{Faculty of Science, Humanities and Education, Technical University of Liberec,  
Studentsk\'{a} 1402/2, CZ-461\,17~Liberec, Czech Republic}
\author{Vladim\'{i}r Karas}
\affiliation{Astronomical Institute, Czech Academy of Sciences, Bo\v{c}n\'{i} II, CZ-141\,00~Prague, Czech~Republic}

\begin{abstract}
Strong gravity and magnetic fields are key ingredients that power processes of accretion and ejection near compact objects. While the particular mechanisms that operate here are still discussed, it seems that the presence of an ordered magnetic field is crucial for the acceleration and collimation of relativistic jets of electrically charged particles on superhorizon length scales. In this context, we further study the effect of a large-scale magnetic field on the dynamics of charged particles near a rotating black hole. We consider a scenario in which the initially neutral particles on regular geodesic orbits in the equatorial plane are destabilized by a charging process (e.g., by photoionization). Some charged particles are accelerated out of the equatorial plane, and they follow jetlike trajectories with relativistic velocities. In our previous paper, we investigated this scenario for the case of perfect alignment of the magnetic field with the axis of rotation; i.e., the system was considered axisymmetric. Here we relax this assumption and investigate nonaxisymmetric systems in which the magnetic field is arbitrarily inclined with respect to the black hole spin. We study the system numerically in order to locate the zones of escaping trajectories and compute the maximum (terminal) escape velocity. It appears that breaking the axial symmetry (even by small inclination angles) substantially increases the fraction of escaping orbits and allows the acceleration to ultrarelativistic velocities that were excluded in the axisymmetric setup. The presence of transient chaotic dynamics in the launching region of  the relativistic outflow is confirmed with chaotic indicators.

\end{abstract}


\section{Introduction}
\label{intro}
The role of large-scale asymptotically uniform magnetic fields in the formation of ``escape zones'' of electrically charged particles near a magnetized rotating black hole was investigated in our previous paper \citep[][hereafter Paper~I]{kopacek18}. These escape zones represent corridors in the phase space along which the particles can be accelerated to large distance and high velocity. We considered the axisymmetric case of perfect alignment of the magnetic field with the axis of rotation. It was found that a certain fraction of the particle population becomes accelerated and ejected out of the system along energetically unbound trajectories. However, only moderate Lorentz factors could be achieved in the axisymmetric configuration. 

Here we perform an analogous study of escaping orbits in a generalized system of an oblique black hole magnetosphere, and we consider the case of an arbitrary inclination between the rotation axis and the asymptotic direction of the magnetic field. We show that inclination of the magnetic field may greatly increase the fraction of escaping particles, and, in particular, the value of the terminal Lorentz factor of the accelerated particles is also enhanced substantially. 

The acceleration of astrophysical outflows and jets from the vicinity of an accreting black hole is supposed to be powered by its rotational energy extracted via the processes of Blandford--Znajek or Blandford--Payne \citep{blandford77,blandford82}, which involve large-scale magnetic fields threading the black hole and the accretion disk as a key ingredient. The plausibility of this scenario is supported by observations and simulations \citep[e.g., ][]{penna10,penna13,liska18,blandford19}. In particular, self-consistent general relativistic magnetohydrodynamic (GRMHD) simulations of accretion processes lead to the emergence of ordered large-scale magnetic fields under rather general conditions \citep[e.g.,][]{tchekhovskoy15,sadowski16}. They typically occur in the diluted region along the black hole axis, while the turbulent small-scale field develops within the accretion torus, where it generates viscosity. Moreover, global structures of ordered magnetic fields also arise in general relativistic kinetic simulations of black hole magnetospheres and jet launching \citep{parfrey19}. First-principle plasma simulations thus further confirm the crucial role of ordered magnetic fields for the acceleration and collimation of the jet.

While the assumption of perfect axisymmetry of the accreting system (as also employed in Paper~I) significantly simplifies analytic calculations and numerical simulations, it is hardly a realistic property of the whole system, as the gas falling from large distances only becomes aware of the direction of the black hole spin when it approaches sufficiently close. The GRMHD simulations of tilted thick accretion disks indeed reveal a ``magneto-spin alignment" mechanism causing a magnetized disk and jet to align with the spin near a black hole while reorienting with the outer disk farther away \citep{mckinney13,liska18}. On the other hand, the inner parts of the thin accretion disk are supposed to align with the spin due to the Bardeen--Petterson effect \citep{bardeen75} caused by Lense--Thirring forces induced by frame dragging. While theorized several decades ago, the effect is still challenging to resolve in simulations. However, it has been recently confirmed for tilted disks of moderate \citep{liska19} to high \citep{liska19b} initial inclinations. In particular, the former paper shows the disk with aspect ratio $H/R\approx 0.03$ and initial tilt of $10\degree$ to warp into alignment with the black hole inside $\approx5\,r_{g}$, where $r_{g}$ is the gravitational radius. Based on these results, only a small misalignment in the form of a nonaxisymmetric perturbation would be tolerated in the inner part of the disk, while a large tilt could be encountered in the outer parts. 

In this context, we investigate the motion of charged particles in the nonaxisymmetric system of a magnetized rotating black hole with the spin and magnetic field misaligned with respect to each other. The present paper is a  direct follow-up to Paper~I, where the alignment was assumed, and we discussed under which circumstances the particles on stable orbits in the equatorial plane may be liberated from the attraction of the center and even accelerated to relativistic velocities due to a sudden charging process occurring on some radius $r_0$. Here we nontrivially enrich our previous discussion by considering an arbitrary inclination between the spin and asymptotic direction of the magnetic field. In particular, we study how the inclination affects the formation of escape zones and the overall effectivity of the acceleration mechanism quantified by the final Lorentz factor of escaping particles. As previously, the system is treated in the single-particle limit, i.e. the particles are noninteracting and the hydrodynamical terms are neglected. The model is thus relevant for the regions with diluted gas where the mean free path of the particles is larger than the characteristic length scale given by the gravitational radius, $r_g=GM/c^2$, where $M$ is the mass of the black hole. The actual physics of accretion onto black holes \citep[e.g., ][]{falanga15} is considerably more complex than the adopted model. In particular, we reduce the interplay between the charged matter and the background magnetic field to the effect of the Lorentz force acting on individual particles, while the more consistent treatment of mutual interaction will be necessary in the dense environment near compact object \citep[e.g., by solving the equations of][]{blandford77}. Moreover, we suppose that matter remains neutral until the ionization process occurring at given radius $r_0$, which may be very close to the horizon where many accretion models expect the matter to be already highly ionized \citep[e.g., ][]{abramowicz13,shakura18}. Due to these limitations, the results of our analysis are mostly of the theoretical interest. The main purpose of the current paper is to develop a toy model to investigate the implications of the nonaxisymmetric perturbation for the stability of bound orbits and for the acceleration of the escaping ones in the magnetosphere shaped by the spinning black hole.

The combined effects of nonaxisymmetry and rotation on the structure of vacuum electromagnetic fields near compact objects has already been investigated by many authors; for the analyses especially close to our context, see \citet{kopacek18b} and \citet{karas13, karas14}. As a consequence of profound changes in the topology of the field induced by breaking the axial symmetry, the dynamics of charged particles also changes dramatically and generally becomes more prone to deterministic chaos. Even a small nonaxisymmetric perturbation may introduce strong chaos in the dynamic system \citep{kopacek14}. In Paper~I, we conjectured that transient chaos plays an important role in launching the outflow in the adopted model. Here we further explore this conjecture and investigate the dynamic regime of escaping particles calculating several chaotic indicators. Namely, we employ the Minkowski--Bouligand box-counting dimension \citep[e.g., ][]{falconer03} and several measures based on the recurrence analysis \citep[][]{marwan07}. 

Escaping orbits in the vacuum magnetospheres of compact objects have already been investigated in several studies. Previously considered setups include a weakly magnetized Schwarzschild black hole \citep{frolov10,alzahrani13}, a strongly magnetized Ernst's spacetime \citep{karas92,huang15}, a spinning black hole with a magnetic test field \citep{alzahrani14,hussain14,shiose14,stuchlik16}, and magnetized naked singularities \citep{babar16}. Most of these studies considered initially bound circular orbits of charged particles destabilized by the kick that was realized as a sudden introduction of the velocity component perpendicular to the orbital plane. We consider, instead, the ionization process of initially neutral particles as an actual trigger of instability. As the initial setup of the model investigated in the present paper is analogous to that of Paper~I, we refer to the more complete introduction and references presented therein (especially see the Section~1 and discussion in Section~4).

The paper is organized as follows. In Section~\ref{spec} we describe the employed model of the oblique black hole magnetosphere and review the equations of motion and the effective potential for charged particles. The trajectories of initially neutral particles escaping from the circular orbits in the equatorial plane are analyzed in Section~\ref{outflow}. The method of the effective potential is used to obtain the necessary conditions for the escape analytically in Section~\ref{ionization}. The emergence and evolution of escape zones with respect to the inclination angle $\alpha$ is then discussed numerically (Section~\ref{escape_zone}).  Acceleration of escaping particles and the maximal Lorentz factor attained within the escape zones are computed in Section~\ref{acceleration}. The role of chaos in the dynamics of escape zones is assessed in Section~\ref{chaos} employing the escape-boundary plots and the chaotic indicators in Section~\ref{rp}. Results are summarized and briefly discussed in Section~\ref{conclusions}. Details of the employed integration scheme are given in the Appendix.

\newpage
\section{Specification of the model, equations of motion}
\label{spec}
The Kerr metric describing the geometry of the spacetime around the rotating black hole is expressed in Boyer--Lindquist coordinates $x^{\mu}= (t,\:r, \:\theta,\:\varphi)$ as follows \citep{mtw}:
\begin{eqnarray}
\label{metric}
{\rm d}s^2&=&-\frac{\Delta}{\Sigma}\Big[{\rm d}t-a\sin{\theta}\,{\rm d}\varphi\Big]^2\\& &\nonumber+\frac{\sin^2{\theta}}{\Sigma}\Big[\big(r^2+a^2)\,{\rm d}\varphi-a\,{\rm d}t\Big]^2+\frac{\Sigma}{\Delta}\,{\rm d}r^2+\Sigma\, {\rm d}\theta^2,
\end{eqnarray}
where
\begin{equation}
{\Delta}\equiv{}r^2-2Mr+a^2,\;\;\;
\Sigma\equiv{}r^2+a^2\cos^2\theta.
\end{equation}
The coordinate singularity at $\Delta=0$ corresponds to the outer/inner horizon of the black hole, $r_\pm=M\pm\sqrt{M^2-a^2}$. Rotation of the black hole is measured by the spin parameter $a\in\left<-M,M\right>$. Here we only consider $a\geq0$ without the loss of generality.

We note that geometrized units are used throughout the paper. The values of basic constants (gravitational constant $G$, speed of light $c$, Boltzmann constant $k$, and Coulomb constant $k_C$) therefore equal unity, $G=c=k=k_C=1$.

We employ the test-field solution of Maxwell's equations for a weakly magnetized Kerr black hole immersed in an asymptotically uniform magnetic field specified by the component $B_z$ parallel to the spin axis and the perpendicular component $B_x$. The electromagnetic vector potential $A_{\mu}$ is given as follows \citep{bicak85}:  

\begin{eqnarray}
\label{empot1}
A_t&=&\frac{B_{z}aMr}{\Sigma}\left(1+\cos^2\theta\right)-B_{z}a\\& &\nonumber+\frac{B_xaM\sin\theta\cos\theta}{\Sigma}\left(r\cos\psi-a\sin\psi\right),\\
A_r&=&-B_x(r-M)\cos\theta\sin\theta\sin\psi\label{empot2},\\
A_{\theta}&=&-B_xa(r\sin^2\theta+M\cos^2\theta)\cos\psi\\& &\nonumber-B_x(r^2\cos^2\theta-Mr\cos2\theta+a^2\cos2\theta)\sin\psi,\label{empot3}\\
A_{\varphi}&=&B_z\sin^2\theta\left[\frac{1}{2}(r^2+a^2)-\frac{a^2Mr}{\Sigma}(1+\cos^2\theta)\right]\label{empot4}\\& &\nonumber-B_x\sin\theta\cos\theta\Big[\Delta\cos\psi\\& &\nonumber+\frac{(r^2+a^2)M}{\Sigma}\left(r\cos\psi-a\sin\psi\right)\Big],
\end{eqnarray}
where $\psi$ denotes the azimuthal coordinate of the Kerr ingoing coordinates, which is expressed in Boyer--Lindquist coordinates as follows:
\begin{equation}
\label{kicpsi}
\psi=\varphi+\frac{a}{r_{+}-r_{-}}\ln{\frac{r-r_{+}}{r-r_{-}}}.
\end{equation}
We notice that $\lim_{r\to \infty}\psi=\varphi$. Setting $B_x=0$ reduces the above vector potential $A_{\mu}$ to the axisymmetric solution \citep{wald74} employed in Paper~I. 

The Hamiltonian $\mathcal{H}$ of a particle of electric charge $q$ and rest mass $m$ in the field $A_{\mu}$ and metric with contravariant components $g^{\mu\nu}$ may be defined as \citep{mtw}
\begin{equation}
\label{hamiltonian}
\mathcal{H}=\textstyle{\frac{1}{2}}g^{\mu\nu}(\pi_{\mu}-qA_{\mu})(\pi_{\nu}-qA_{\nu}),
\end{equation}
where $\pi_{\mu}$ is the generalized (canonical) momentum. The equations of motion are expressed as
\begin{equation}
\label{hameq}
\frac{{\rm d}x^{\mu}}{{\rm d}\lambda}\equiv p^{\mu}=
\frac{\partial \mathcal{H}}{\partial \pi_{\mu}},
\quad 
\frac{d\pi_{\mu}}{d\lambda}=-\frac{\partial\mathcal{H}}{\partial x^{\mu}},
\end{equation}
where $\lambda\equiv\tau/m$ is a dimensionless affine parameter ($\tau$ denotes the
proper time). Employing the first equation, we obtain the kinematical four-momentum as $p^{\mu}=\pi^{\mu}-qA^{\mu}$, and the conserved value of the Hamiltonian is therefore given as $\mathcal{H}=-m^2/2$. The system is stationary, and the time component of canonical momentum $\pi_t$ is therefore an integral of motion that equals the (negatively taken) energy of the test particle $\pi_t\equiv-E$. In the rest of the paper, we switch to the specific quantities $E/m\rightarrow E$, $q/m\rightarrow q$, which corresponds to setting the rest mass of the particle $m=1$ in the formulae.

An effective potential expressing the minimal allowed energy of charged test particles in a nonaxisymmetric magnetosphere of a rotating black hole may be derived in the rest frame of a static observer \citep{kopacek18c}. The tetrad vectors of this frame are given as \citep{semerak93}

\begin{eqnarray}
\label{statictetrad1}
e_{(t)}^{\mu}&=& \left[\frac{\Sigma^{1/2}}{\chi},0,0,0\right],\;\;\;
e_{(r)}^{\mu}=\left[0,\frac{\Delta^{1/2}}{\Sigma^{1/2}},0,0\right],\\
\label{statictetrad2}
e_{(\theta)}^{\mu}&=&\left[0,0,\frac{1}{\Sigma^{1/2}},0\right],\;\;\;\\
\label{statictetrad3}
e_{(\varphi)}^{\mu}&=&\frac{\chi}{\sin\theta\Delta^{1/2}\Sigma^{1/2}}\left[\frac{-2aMr\sin^2\theta}{\chi^2},0,0, 1\right],
\end{eqnarray}
where $\chi^2\equiv\Delta-a^2\sin^2\theta$.

The static frame is employed to express the effective potential, 
\begin{equation}
\label{eff_pot}
V_{\rm eff}(r,\theta,\varphi)=\left(-\beta+\sqrt{\beta^2-4\alpha\gamma}\right)/2\alpha,
\end{equation}
with the coefficients defined as
\begin{equation}
\alpha=\left[e^t_{(t)}\right]^2,\;\;\;\beta=2 q A_t e^t_{(t)},\;\;\;\gamma=q^2\left[e^t_{(t)}\right]^2A_t^2-1.
\end{equation}

Kerr spacetime allows no static observers inside the ergosphere, whose boundary (corresponding to $\chi^2=0$) is given as $r_{\rm{s}}=M+\sqrt{M^2-a^2\cos{\theta}^2}$. However, unlike the axisymmetric case, neither nonstatic frames with $u^{\varphi}\neq 0$, nor the Boyer--Lindquist coordinate basis itself may be used to express the effective potential in the case of the oblique magnetosphere \citep{kopacek18c,kopacek14}, and the potential (\ref{eff_pot}) is thus well defined outside the ergosphere only.

\section{Escape of electrically charged particles from the magnetized accretion disk}
\label{outflow}
\subsection{Necessary Conditions for the Escape of Charged Particles}
\label{ionization}
We assume a thin-disk accretion geometry \citep{novikov73} in which the electrically neutral matter gradually descends between corotating (prograde, equatorial) Keplerian orbits with energy $E_{\rm Kep}(r)$ and angular momentum $L_{\rm Kep}(r)$, given as follows \citep{bardeen72}:
\begin{eqnarray}
 \label{kepconst1}
E_{\rm Kep}&=&\frac{r^2-2Mr+ a \sqrt{Mr}}{r\sqrt{r^2-3Mr+ 2a\sqrt{Mr}}},\\
\label{kepconst2}L_{\rm Kep}&=&\frac{\sqrt{M} (r^2+a^2- 2a\sqrt{Mr})}{\sqrt{r(r^2-3Mr+ 2a\sqrt{Mr})}}.
\end{eqnarray}
Circular geodesics are not allowed below the radius of the innermost stable circular orbit (ISCO), whose position for corotating particles is given as
\begin{equation}
\label{rms}
r_{\rm{ISCO}}=M\left(3+Z_2-\sqrt{(3-Z_1)(3+Z_1+2Z_2)}\right),
\end{equation} 
where $Z_1\equiv1+\left(1-\frac{a^2}{M^2}\right)^{1/3}\left[\left(1+\frac{a}{M}\right)^{1/3}+\left(1-\frac{a}{M}\right)^{1/3}\right]$ and $Z_{2} \equiv \sqrt{\frac{3a^2}{M^2}+Z_1^2}$.

Below the ISCO, the geodesics are supposed to turn into freely falling inspirals maintaining the energy and angular momentum corresponding to the ISCO radius; i.e., the particles keep $E=E_{\rm Kep}(r_{\rm ISCO})$ and $L=L_{\rm Kep}(r_{\rm ISCO})$ during their infall.

However, initially neutral elements may undergo a sudden charging process (by photoionization, for example) at a given radius $r_0$ and obtain a nonzero specific charge $q$. While the change of the rest mass $m$ may be neglected, the particle dynamics changes due to the presence of the electromagnetic field $A_{\mu}$. In particular, the conserved value of energy is modified as
\begin{equation}
\label{newenergy}
E=E_{\rm Kep}-qA_t,
\end{equation}
while the values of the spatial components of canonical momentum are changed as
\begin{equation}
\label{newmomenta}
\pi_r=\pi_r^0+qA_r,\;\;\;\pi_{\theta}=qA_{\theta},\;\;\;\pi_{\varphi}=L_{\rm Kep}+qA_{\varphi}, 
\end{equation}
where $\pi_r^0$ is zero for particles ionized above/at the ISCO, and for infalling particles with ionization radius $r_0 < r_{\rm{ISCO}}$, the value is calculated using the assumption that the particle is confined to the equatorial plane ($\pi_{\theta}^0=0$) before it obtains the electric charge.

Once the charge is introduced, the value of the effective potential, given by the Equation~(\ref{eff_pot}), changes accordingly. In order to study the trajectories of escaping particles, we examine the behavior of the potential for $r\gg M$. In particular, for the initially neutral particle with energy $E_{\rm Kep}$ ionized in the equatorial plane at $r_0$, we obtain the following relation valid in the asymptotic region: 

\begin{equation}
E-V_{\rm eff}|_{r\gg M}= E_{\rm Kep}-1-\frac{qB_z a}{r_0}+\mathcal{O}\left(r^{-1}\right).
\label{asym}
\end{equation}

Motion is possible only for $E\geq V_{\rm eff}$. Since $E_{\rm Kep}<1$ with finite $r_0$ and $a \geq 0$ is considered, we observe that (i) particles may only escape for $qB_z<0$, (ii) the escape is possible only for $a\neq 0$, (iii) the asymptotic velocity of escaping particles is an increasing function of parameters $|qB_z|$ and $a$, and a decreasing function of $r_0$, and (iv) the escape is not allowed for the perpendicular inclination, $\alpha \equiv \arctan\left( B_x / B_z \right)=\pi/2$. 

Conditions (i)-(iii) are analogous to those obtained in Paper~I for an aligned case with an asymptotic magnetic field $B=B_z$. Indeed, the perpendicular component $B_x$ does not contribute to the energy of the charged particle in the given setup, since the corresponding terms in the time component of the vector potential (\ref{empot1}) vanish in the equatorial plane and the $B_x$ terms in $V_{\rm eff}$ decrease as $\mathcal{O}\left(r^{-1}\right)$ or faster. Therefore, for a given asymptotic magnitude of the field $B\equiv\sqrt{B_x^2+B_z^2}$, the energy and thus the final escape velocity should decrease with increasing inclination $\alpha$. For the case of the field perpendicular to the rotation axis ($\alpha=\pi/2$), the escape is not allowed at all according to condition (iv). The effective potential for the axisymmetric system defined by Equation~(6) of Paper~I was more specific and also predicted the escape direction along the axis of symmetry. Although the directionality of trajectories cannot be inferred from Equation~ (\ref{asym}), we expect the particles at $r\gg M$ to move along the magnetic field, which is asymptotically uniform in asymptotically flat spacetime.

Analysis of the effective potential shows that escaping orbits are, in principle, possible in the given setup. The first condition, $qB_z < 0$, requires that the charge of the particle is negative for the parallel orientation of the spin and the magnetic field component $B_z$, and it is positive for the antiparallel orientation. The escaping orbits are not allowed in the Schwarzschild limit, $a=0$, regardless of the values of $q$, $B_z$, $B_x$, and $r_0$. The asymptotic velocity of escaping particles rises with increasing values of $a$, $|q|$, and $|B_z|$ and will be higher for the particles charged closer to the horizon. The escape of particles is not allowed in the perpendicular configuration of the field ($B_z =0$) regardless of the values of $B_x$ and other parameters. 

We conclude that all necessary conditions for the escape may be fulfilled in the given setup. However, as we have demonstrated in Paper~I, for the special case of an axisymmetric system with $B_x =0$, these conditions are not sufficient; to test whether particles really escape and to discuss their asymptotic velocity, we need to employ a numerical approach and integrate the equations of motion (\ref{hameq}). In the rest of the paper, we switch to dimensionless units and set $M=1$.

\begin{figure*}[ht]
\center
\includegraphics[scale=.39]{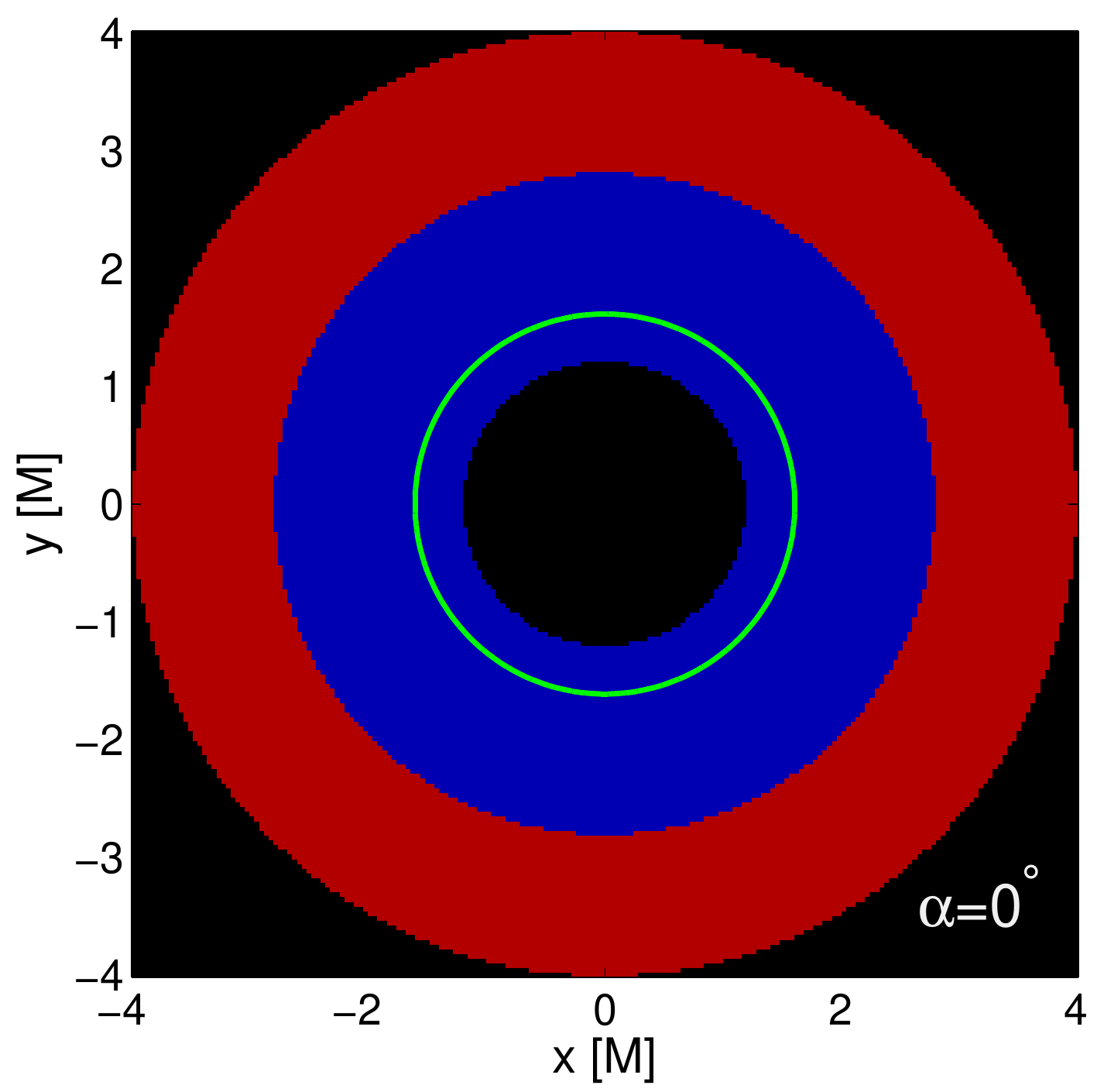}
 \includegraphics[scale=.39]{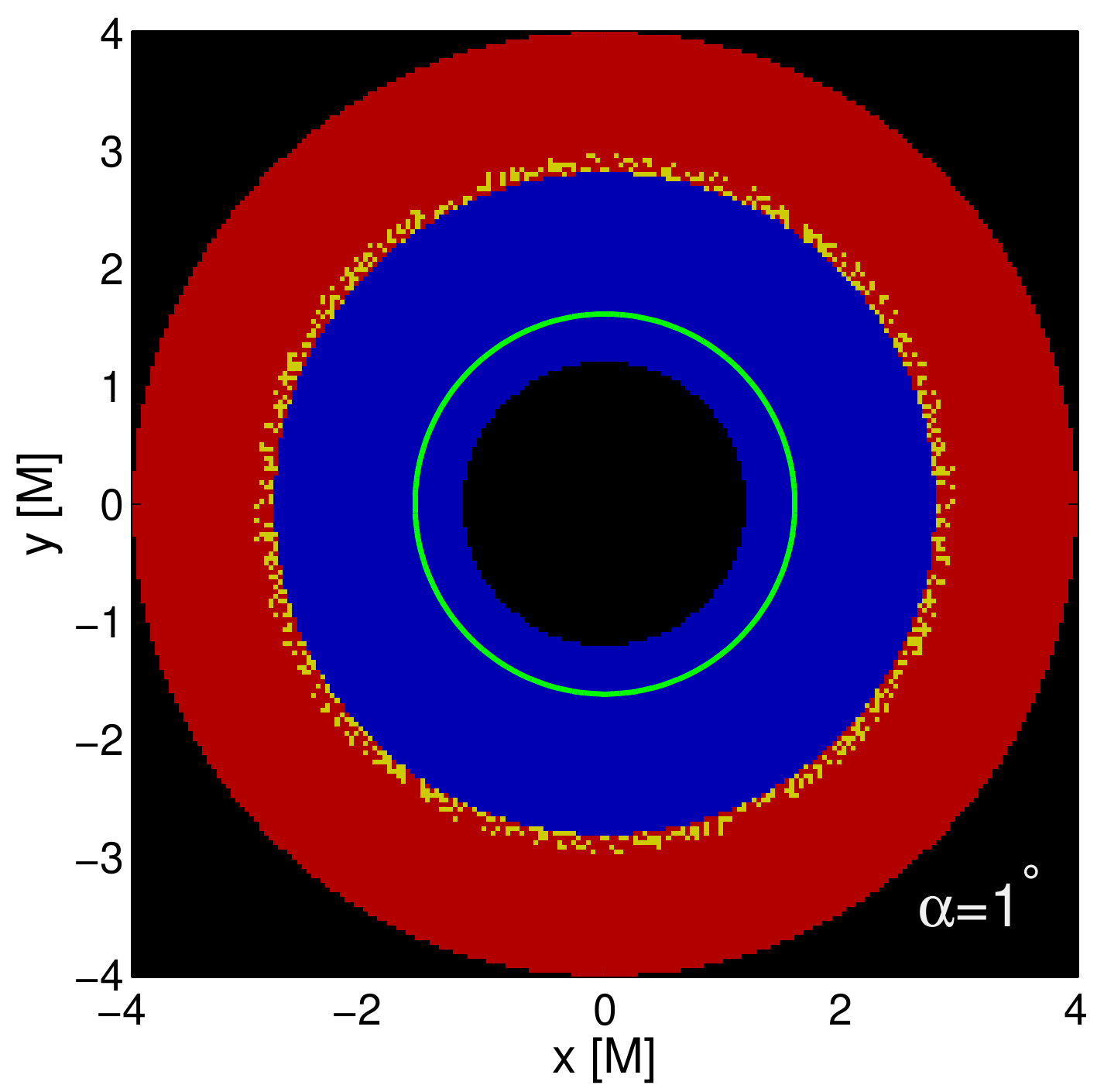}
\includegraphics[scale=.39]{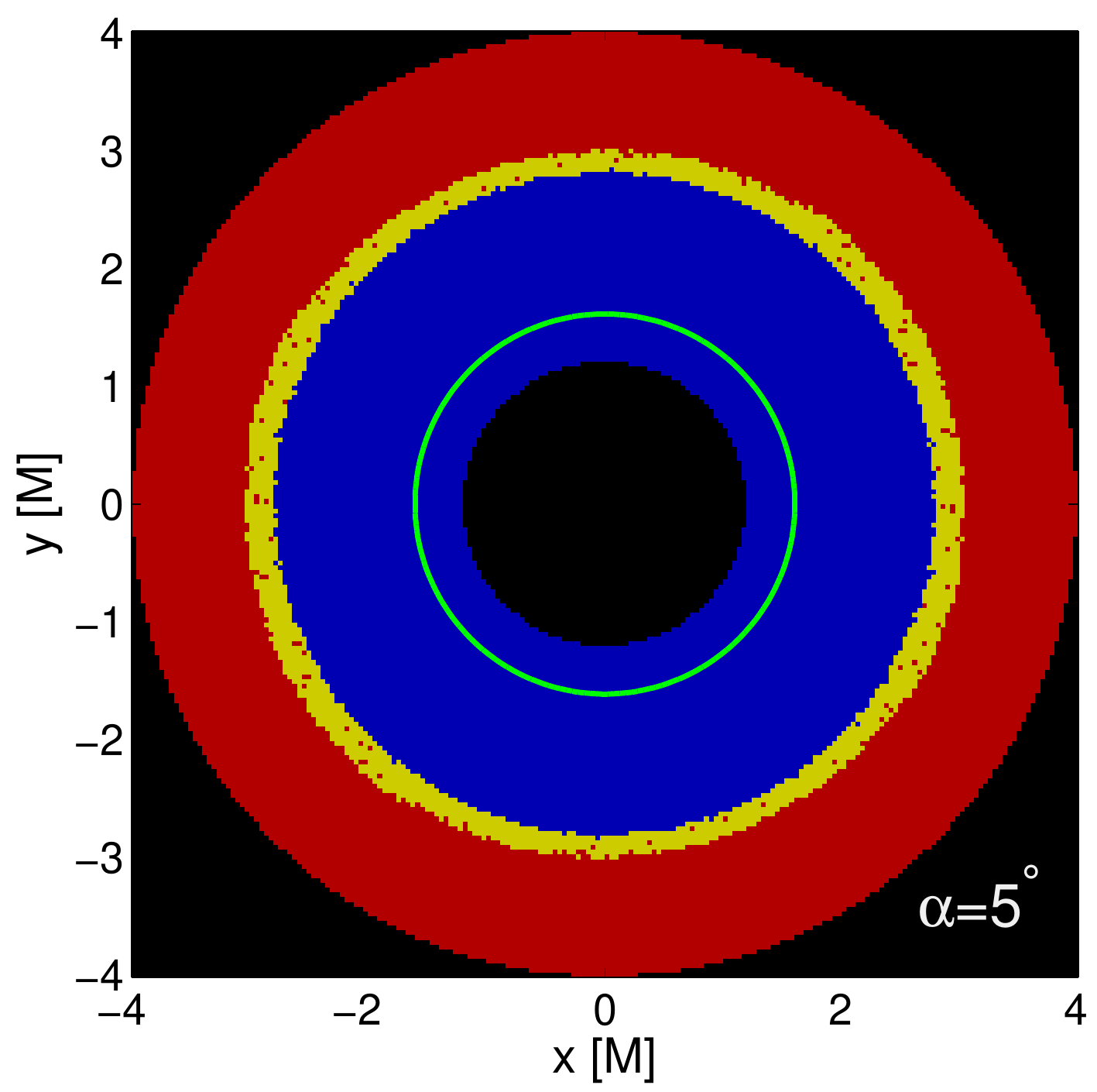}\\
\includegraphics[scale=.39]{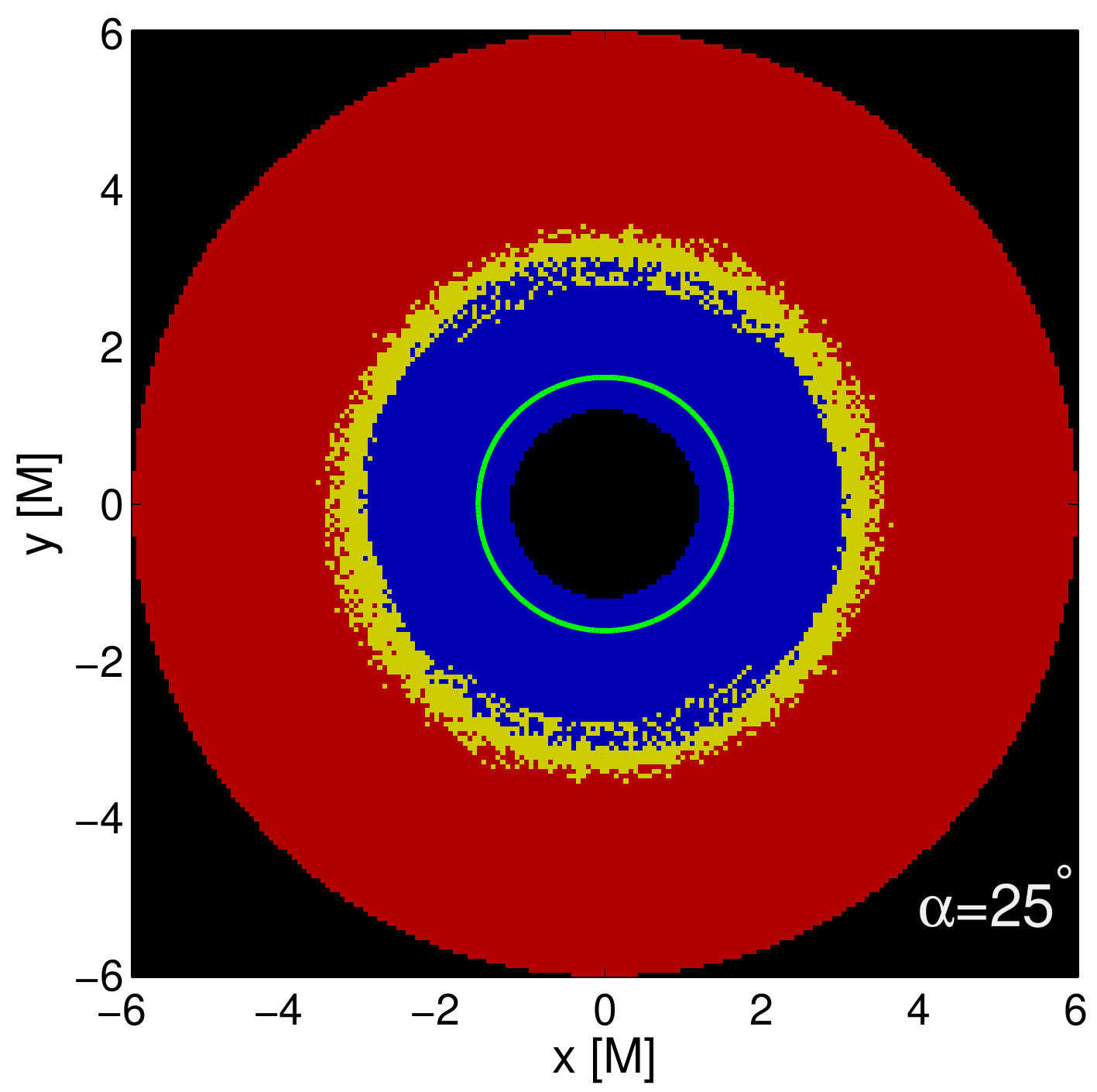}
 \includegraphics[scale=.39]{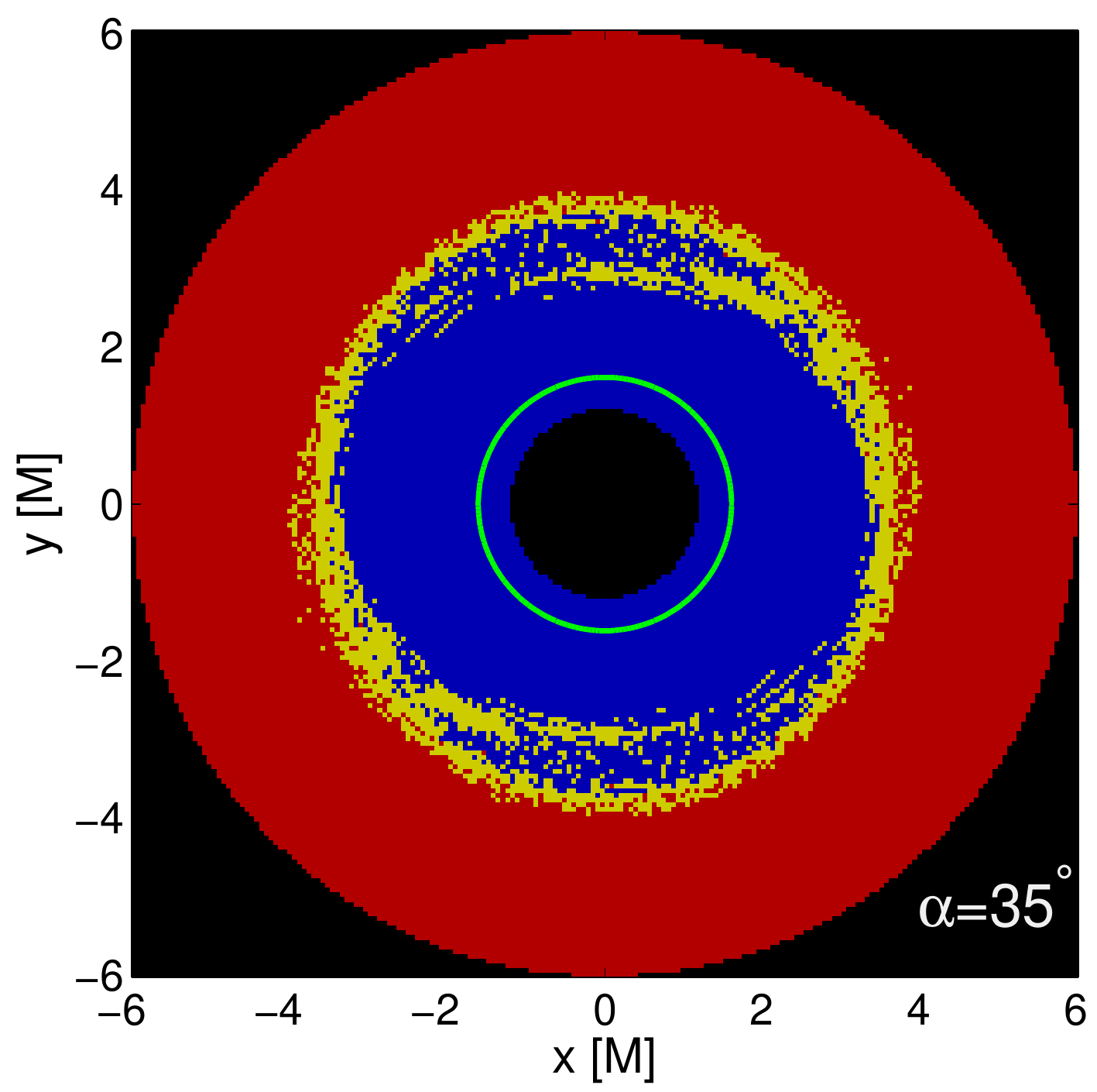}
\includegraphics[scale=.39]{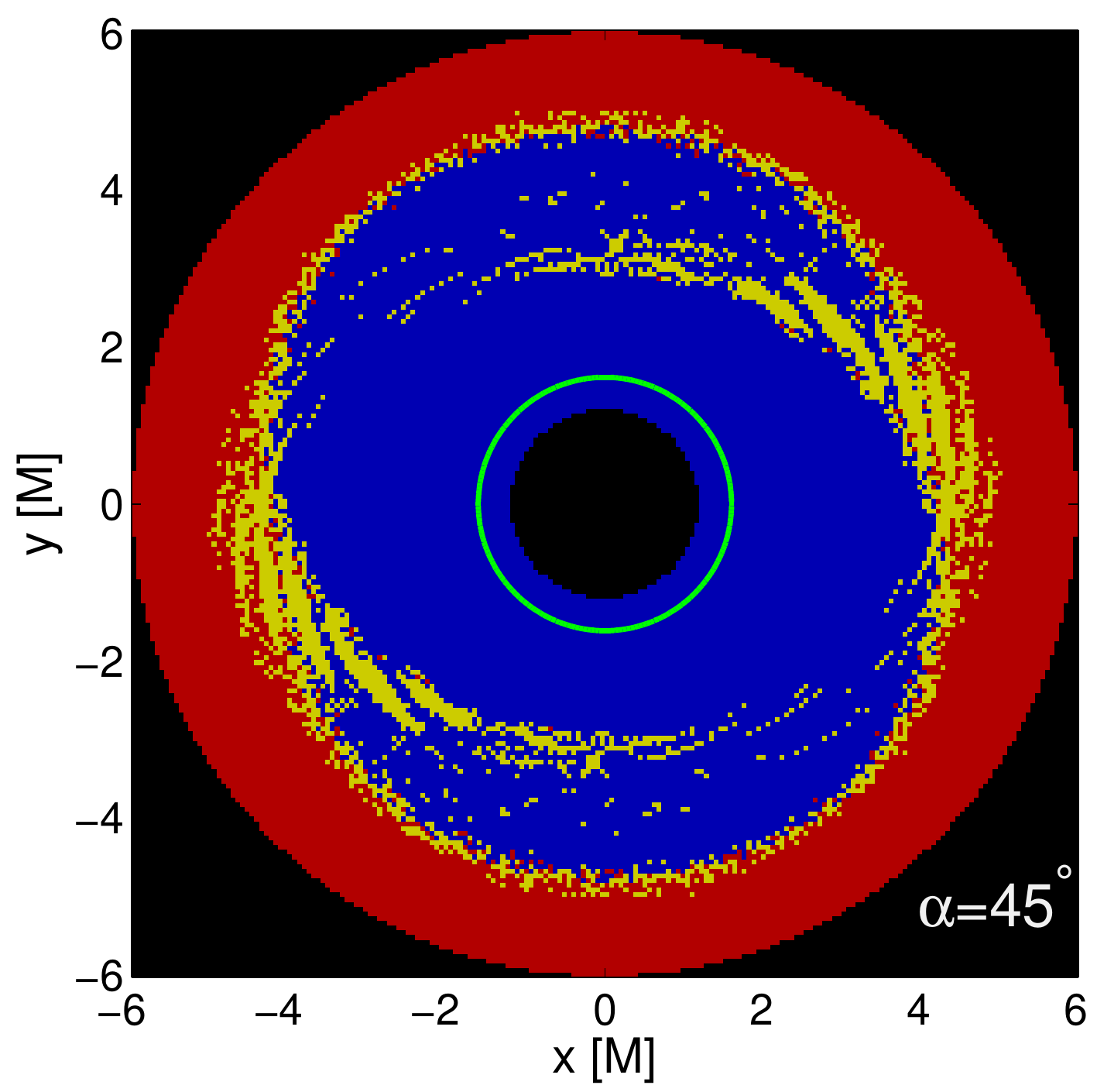}
\caption{Different types of the trajectories launched from the equatorial plane $(x,y)$ are plotted with respect to the magnetic inclination angle $\alpha$ (in degrees). Color-coding: blue for plunging orbits, red for stable ones (bound to the black hole), and yellow for escaping trajectories. The green circle denotes the ISCO. The inner black region marks the horizon of the black hole. The parameters of the system are $a=0.98$ and $qB=-5$. The magnetic field is inclined in the positive $x$-direction.}
\label{escape_primary_zone}
\end{figure*}

\subsection{Escape Zones in Oblique Magnetosphere}
\label{escape_zone}
The {\em escape zones} describe regions where the particles get accelerated to attain the escape velocity from the system. The emergence of escape zones and acceleration of escaping particles in the special case of an aligned magnetic field ($B=B_z$, $B_x=0$) was discussed in Paper~I. A numerical survey revealed that escaping trajectories are realized only in a limited region of parameter space. In particular, escaping trajectories were found only for $qB\lessapprox -0.5$. In the interval $-4.5\lessapprox qB\lessapprox -0.5$, we could observe in the $r_0\times a$ plane (ionization radius vs. spin) a narrow escape zone stretching from extremal spin $a=1$ to lower spin values. However, for $qB\lessapprox-4.5$, it disconnects from the $a=1$ limit, and the whole escape zone is gradually compressed to lower spin and smaller radii as the $|qB|$ increases. For high values of spin, the escaping trajectories were allowed only for a small range of the magnetization parameter $qB$. The resulting acceleration of escaping particles was then limited. One of the key questions we address in the current paper is whether these limitations are also present in the oblique magnetospheres, and, in particular, whether the nonaxisymetric system could support the acceleration to ultrarelativistic velocities.

The inclination of the field breaks the axial symmetry of the system and cancels the conservation of the axial component of the angular momentum. The discussion of particle trajectories is thus enriched by new parameters compared to the axisymmetric case. The magnetic field is now specified by two parameters, namely, components $B_z$ and $B_x$ or, equivalently, the field magnitude $B\equiv \sqrt{B_z^2+B_x^2}$ and inclination angle $\alpha\equiv \arctan\left( B_x / B_z \right)$. The initial position of the particle in the equatorial plane ($\theta_0=\pi/2$) is defined by the ionization radius $r_0$ and azimuthal angle $\varphi_0$. In the considered scenario, it follows from Eqs.~(\ref{empot3}) and (\ref{newmomenta}) that $\pi^0_{\theta}(\varphi+\pi)=-\pi^0_{\theta}(\varphi)$, and trajectories with initial azimuthal angles $\varphi_0$ and $\varphi_0+\pi$ are thus equivalent. 

\begin{figure*}[ht]
\center
\includegraphics[scale=.39]{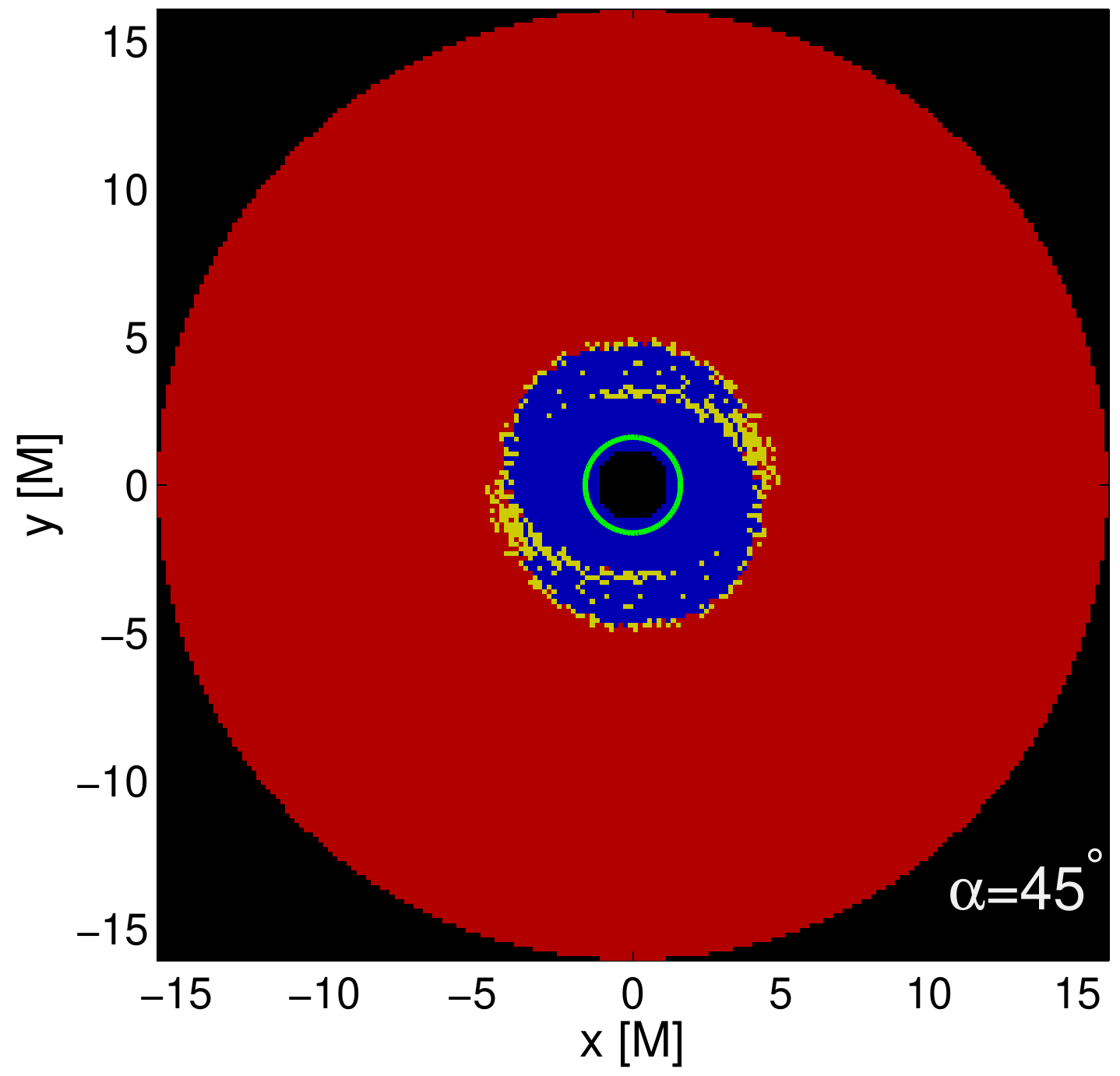}
 \includegraphics[scale=.39]{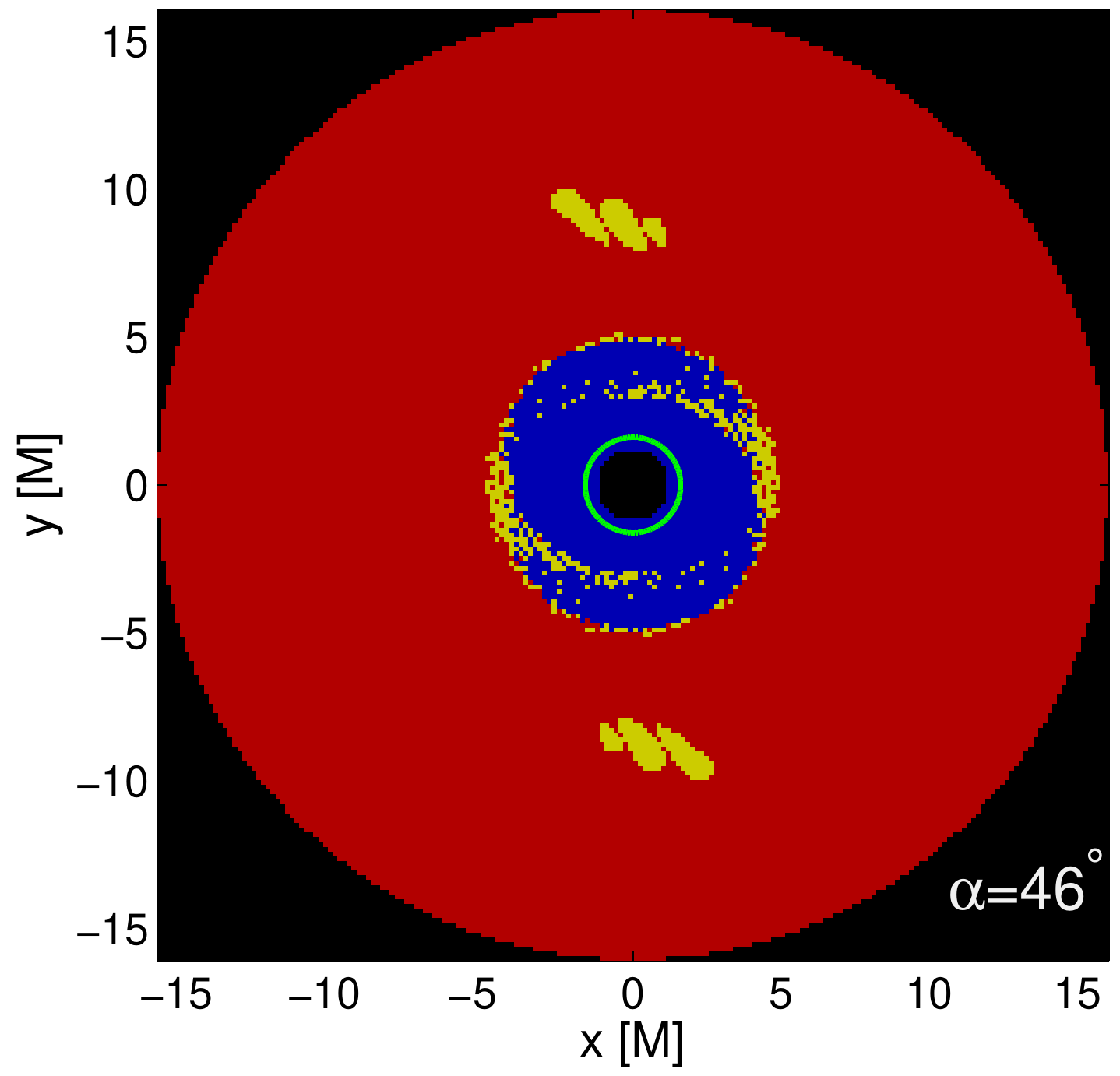}
\includegraphics[scale=.39]{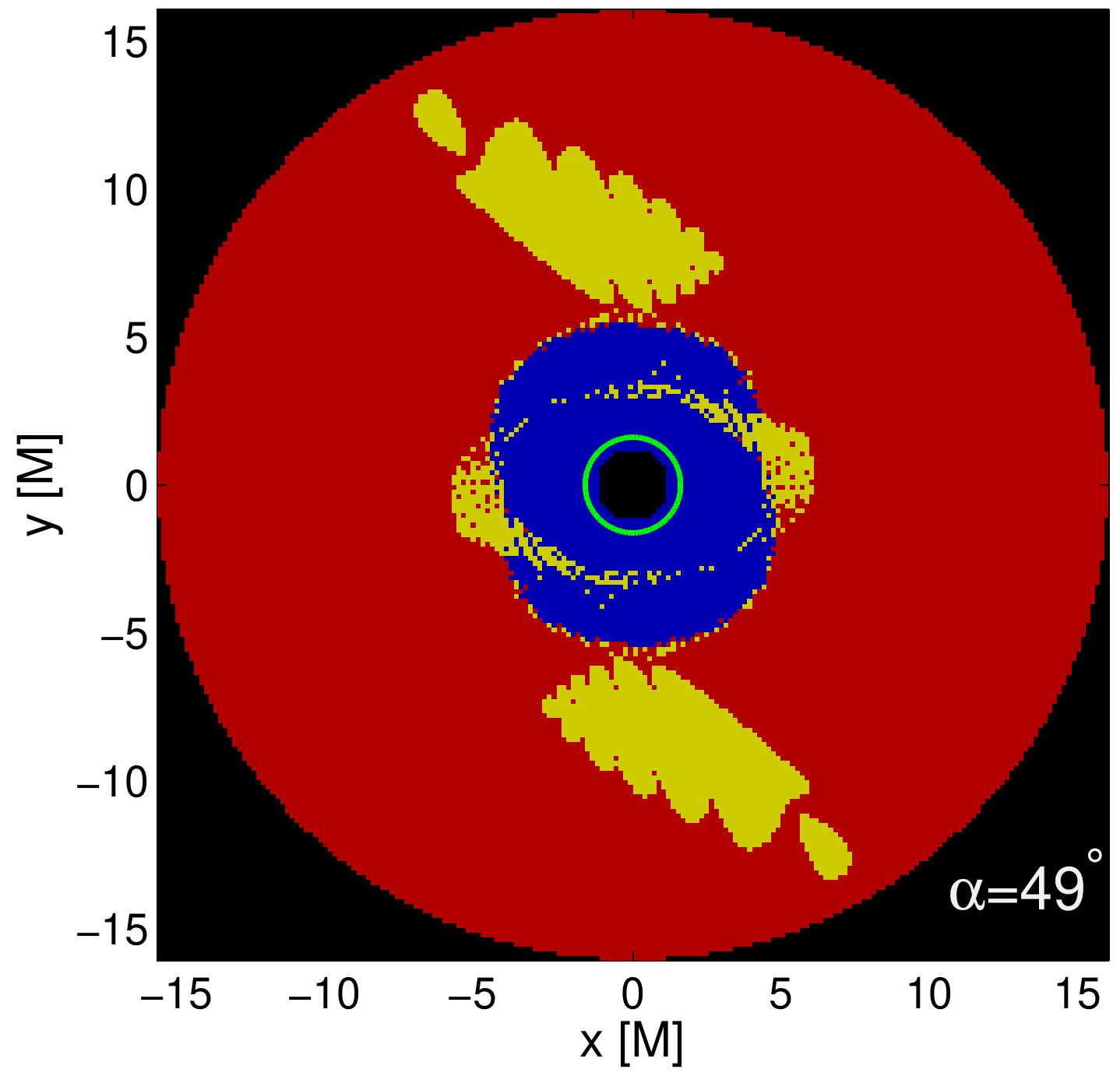}\\
\includegraphics[scale=.39]{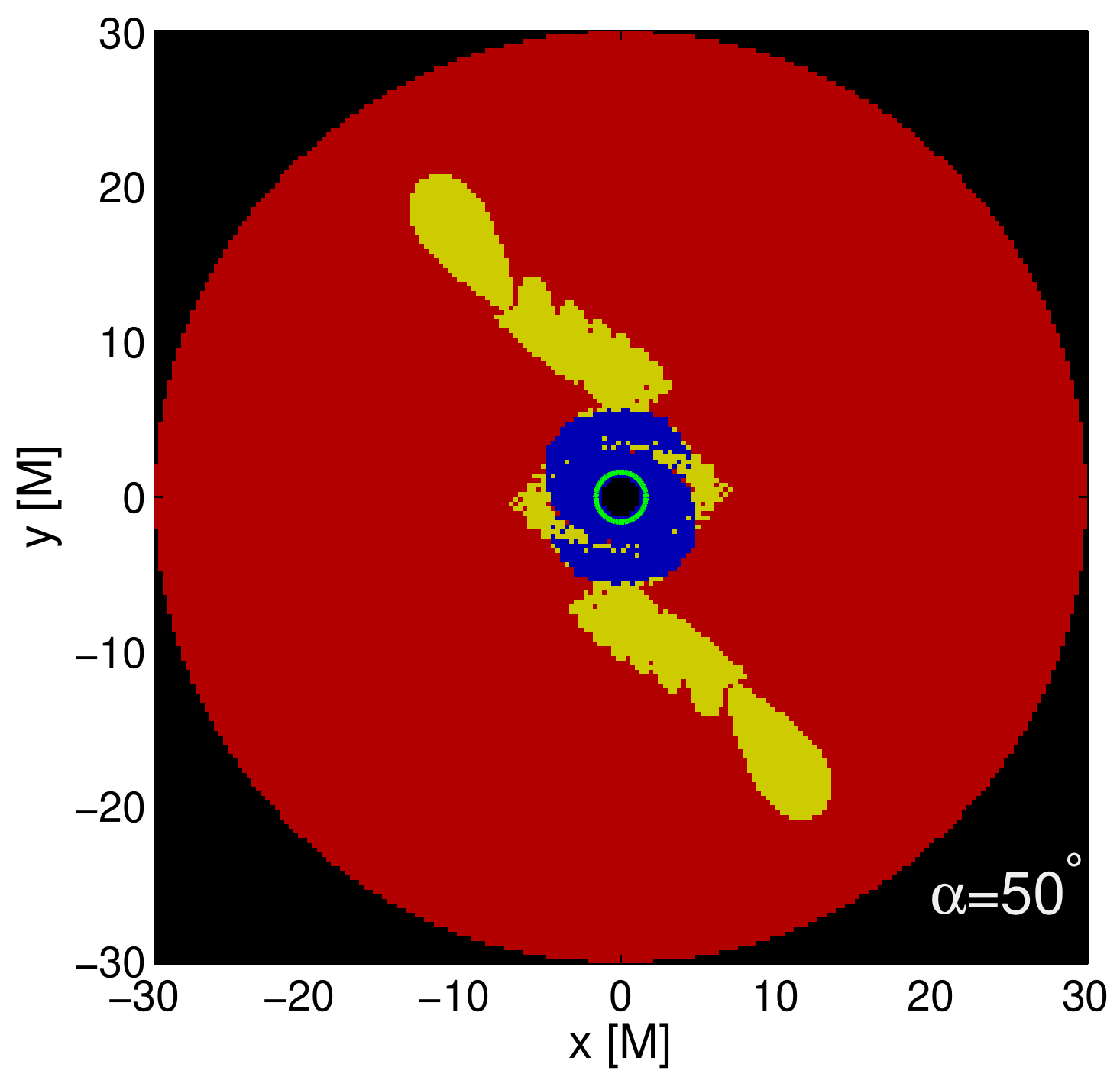}
\includegraphics[scale=.39]{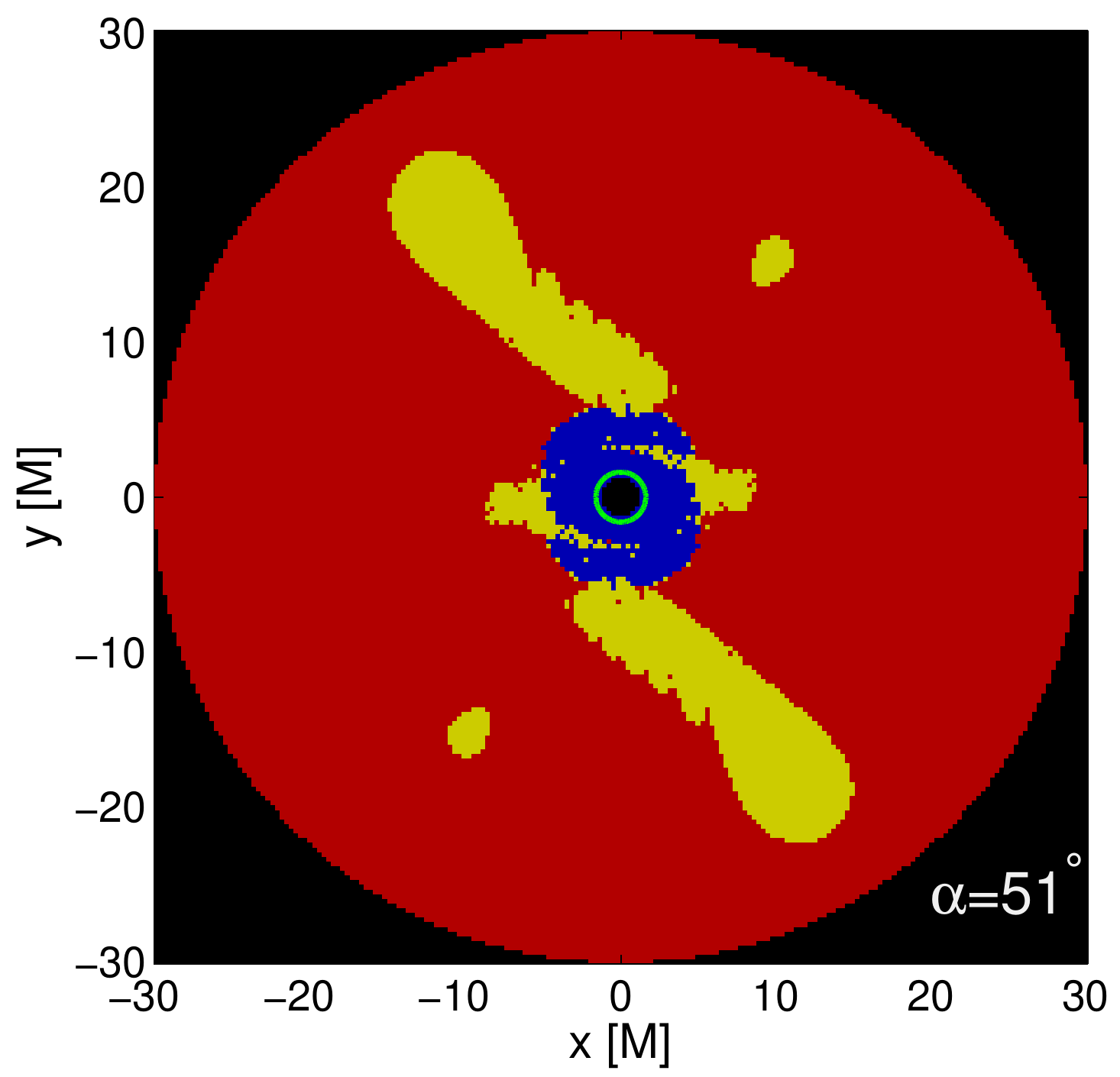}
\includegraphics[scale=.39]{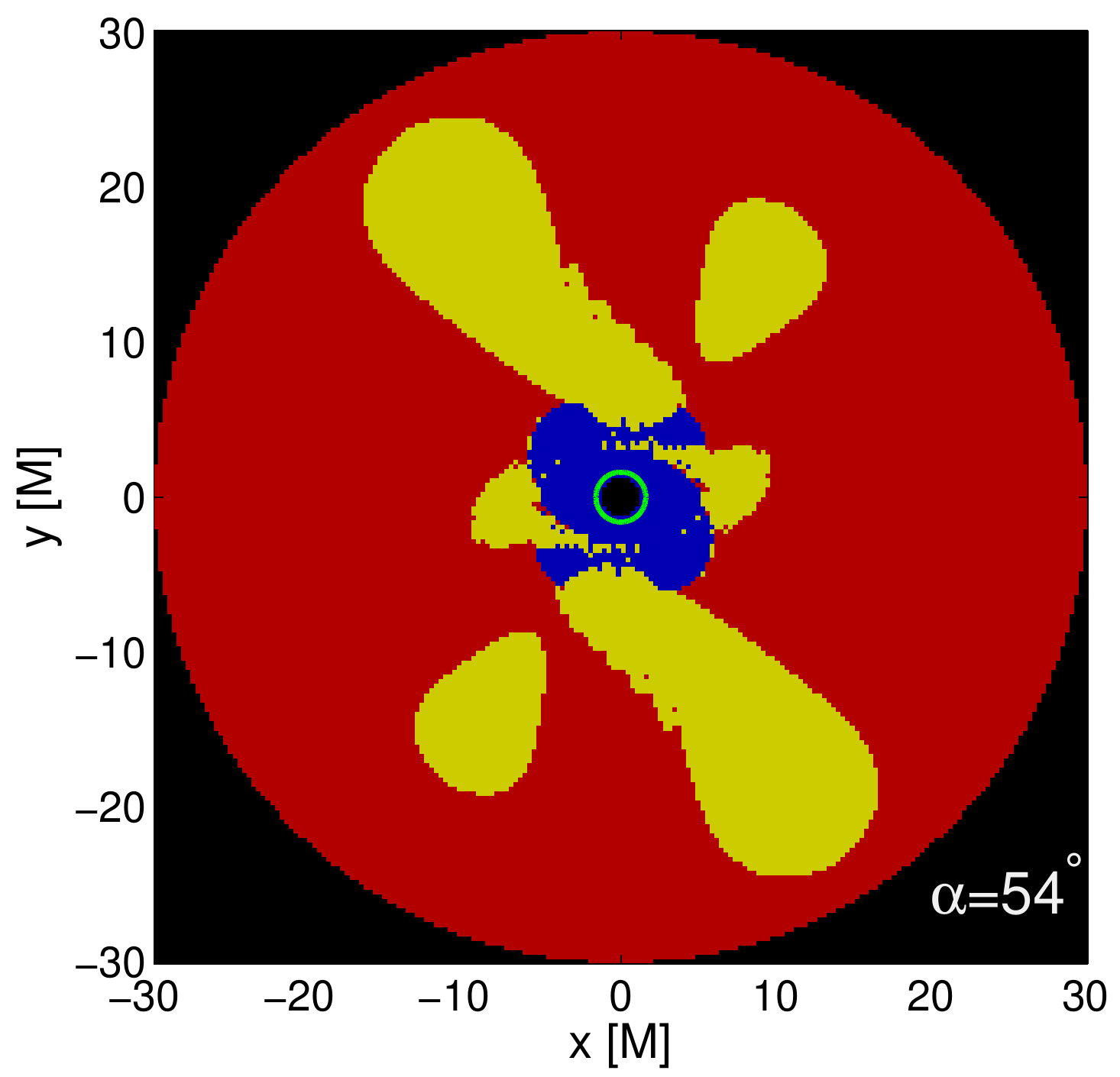}
\caption{Emergence and evolution of the secondary (top panels) and tertiary (bottom panels) escape zones in the equatorial plane with respect to the magnetic inclination angle $\alpha$. Color-coding and parameter choice are the same as in Figure~\ref{escape_primary_zone}.}
\label{escape_23_zone}
\end{figure*}

For a given value of spin $a$, inclination $\alpha$, and magnetization parameter $qB$, we numerically integrate the system of Hamiltonian equations of motion (\ref{hameq}) with the set of initial positions in the equatorial plane (i.e., with particles differing in initial (ionization) radius $r_0$ and $\varphi_0$). If the particle does not plunge into the horizon, we integrate its trajectory up to $\lambda_{\rm fin}=1000$ and consider it escaping if $r_{\rm fin}\geq 200$. In the plots, we use the following color-coding based on the final states: blue dot for the plunging orbits, red for the stable trajectories (with $r_{\rm fin}< 200$), and yellow dot for the escaping particles. The typical resolution of the plots is 200 trajectories per diameter of the investigated portion of the equatorial plane. Integration of non-linear equations (\ref{hameq}) must be performed with caution, as an inappropriate choice of the integration scheme might easily lead to unreliable results. To control the global integration error we evaluate the relative error of energy $E_{rr}\equiv\left|E_n/E-1\right|$, where $E_n$ is an actual value of energy expressed from the normalization condition $p^{\mu}p_{\mu}=-1$, which is affected by the integration errors of the coordinates and momenta, while $E$ denotes its initial value, which is conserved by a real trajectory, since the relevant equation of motion reads $\frac{d\pi_t}{d\lambda}=0$. A numerical error of energy induces the artificial excitation or damping of the system, and since we deal with the nonintegrable system that is sensitive to initial conditions, the value of $E_{rr}$ must be kept within reasonable bounds. Details of the employed integration routine are provided in the Appendix. 

The emergence and evolution of the primary escape zone with respect to the inclination angle $\alpha$ for the particular choice of parameters ($a=0.98$ and $qB=-5$) are shown in Figure~\ref{escape_primary_zone}. We intentionally choose such values of the spin and magnetization parameters for which the escape zone does not form in the aligned field. Indeed, for $\alpha=0^{\degree}$, we observe no escaping orbits in the first panel of Figure~\ref{escape_primary_zone}. However, a slight perturbation of the axial symmetry with the field inclination $\alpha=1^{\degree}$ appears sufficient to trigger the formation of the escape zone. With increasing inclination, the escape zone becomes more pronounced and it extends to higher radii. The azimuthal asymmetry of the zone becomes apparent for higher inclinations, as shown in the bottom panels of Figure~\ref{escape_primary_zone}.

\begin{figure*}[ht]
\center
\includegraphics[scale=.39]{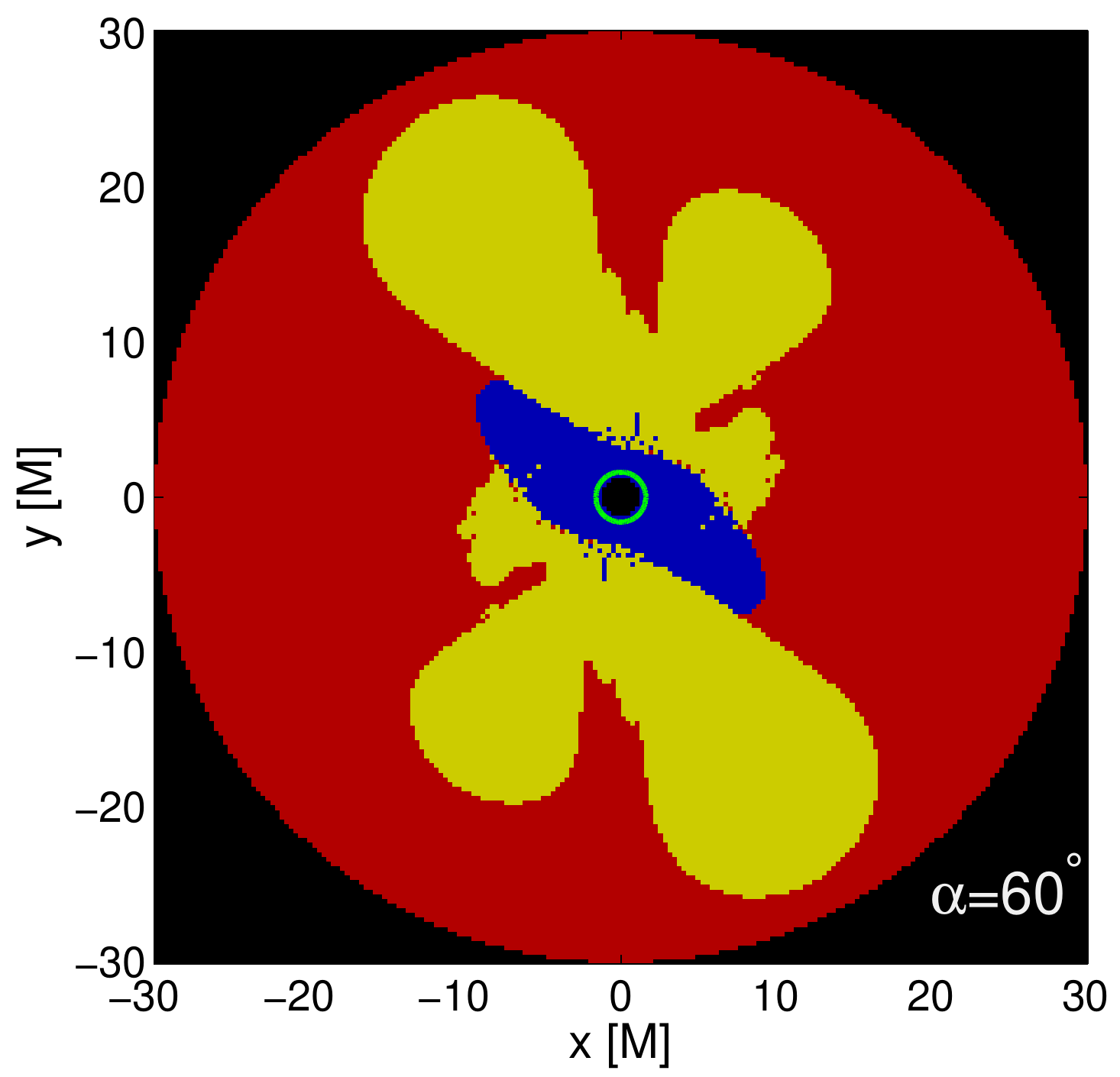}
\includegraphics[scale=.39]{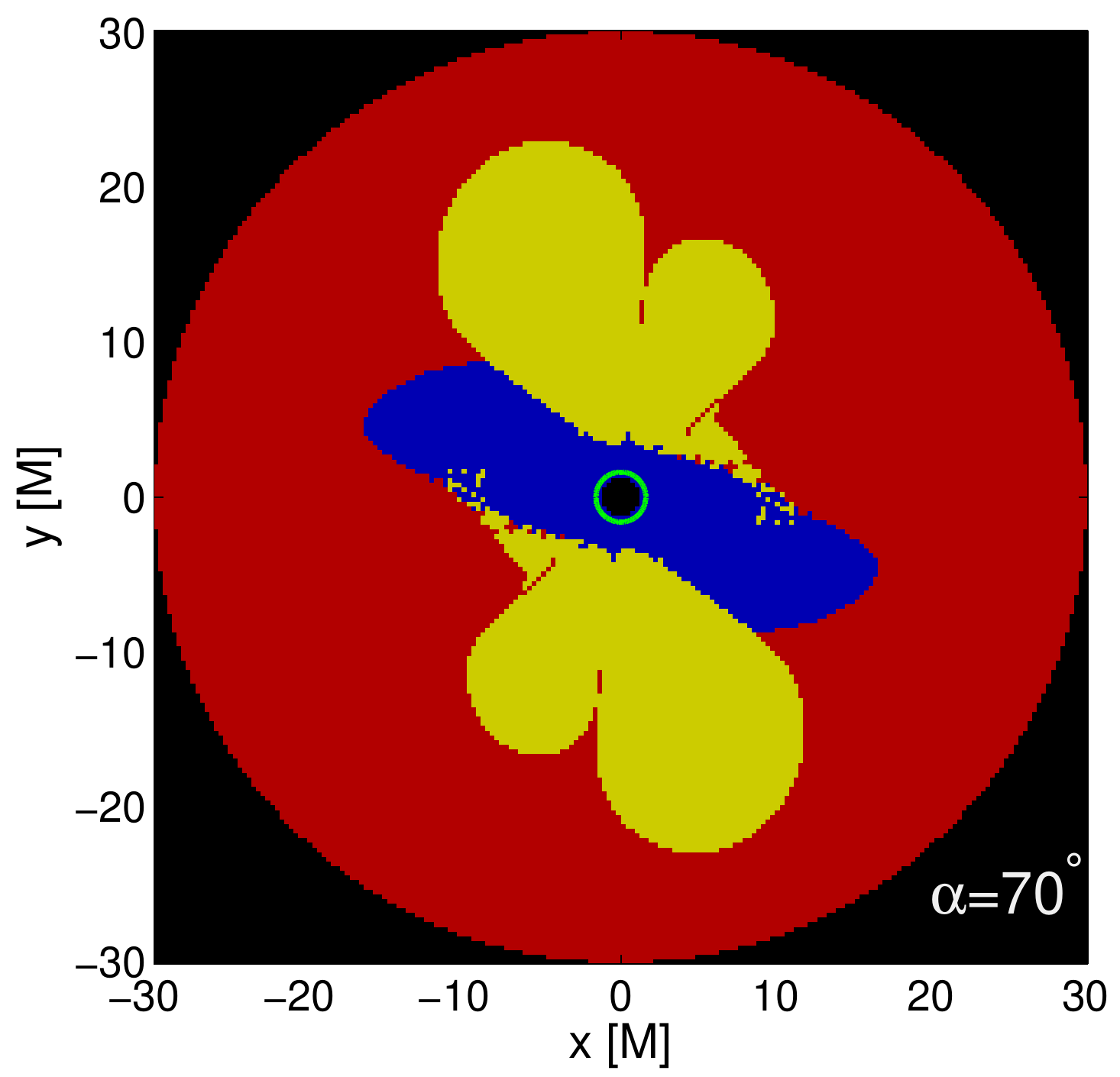}
\includegraphics[scale=.39]{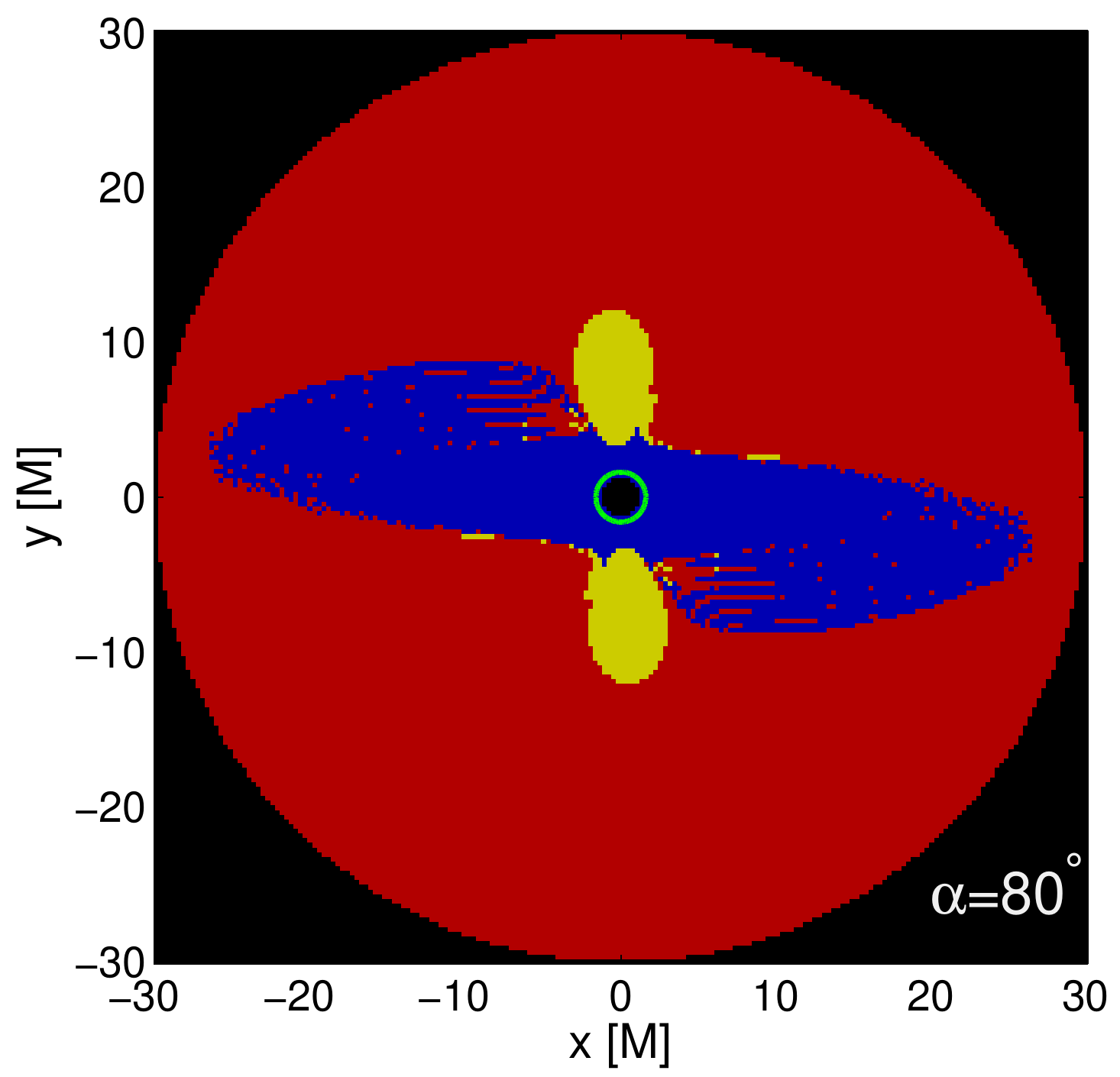}\\
\includegraphics[scale=.39]{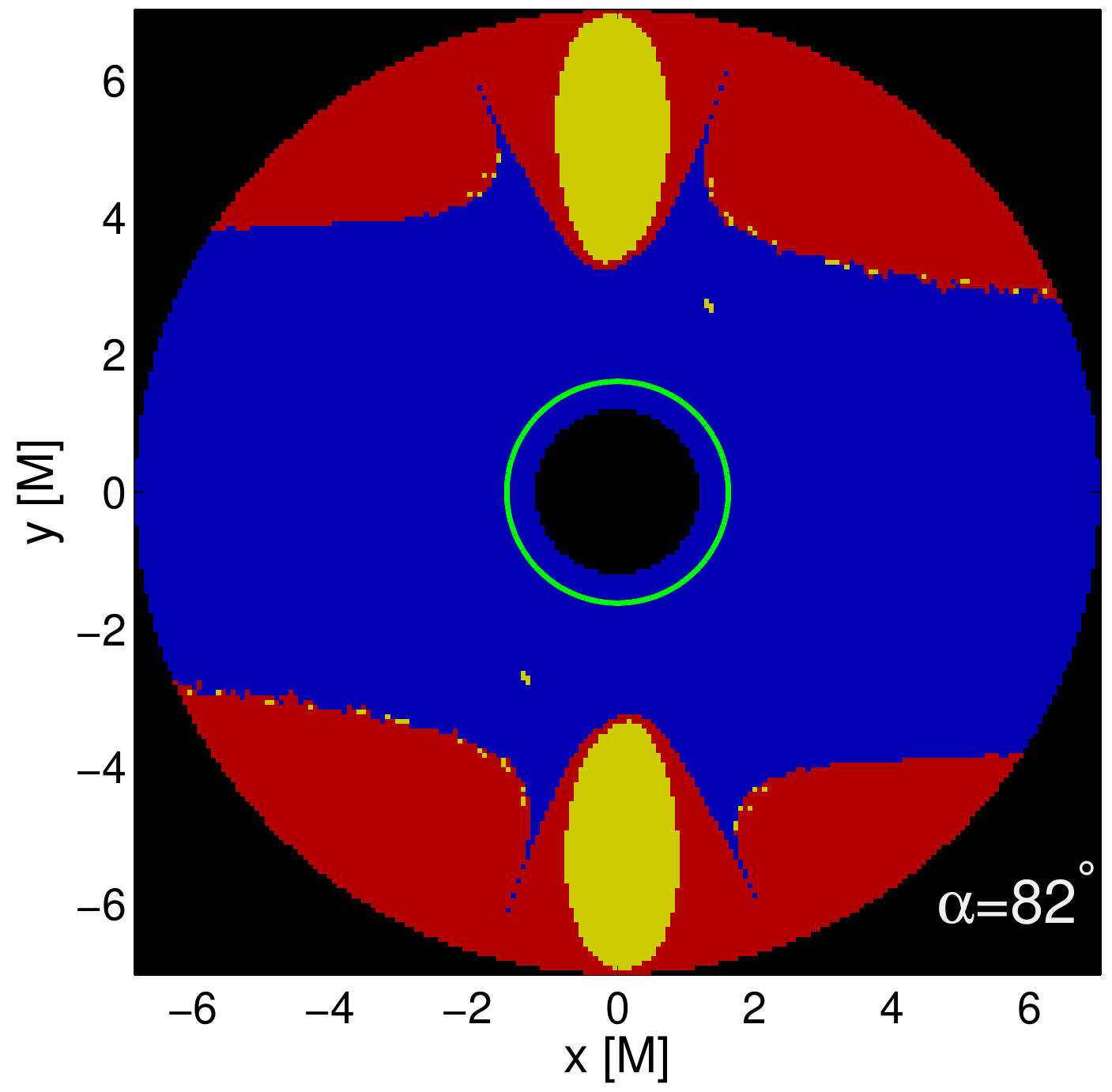}~~
\includegraphics[scale=.39]{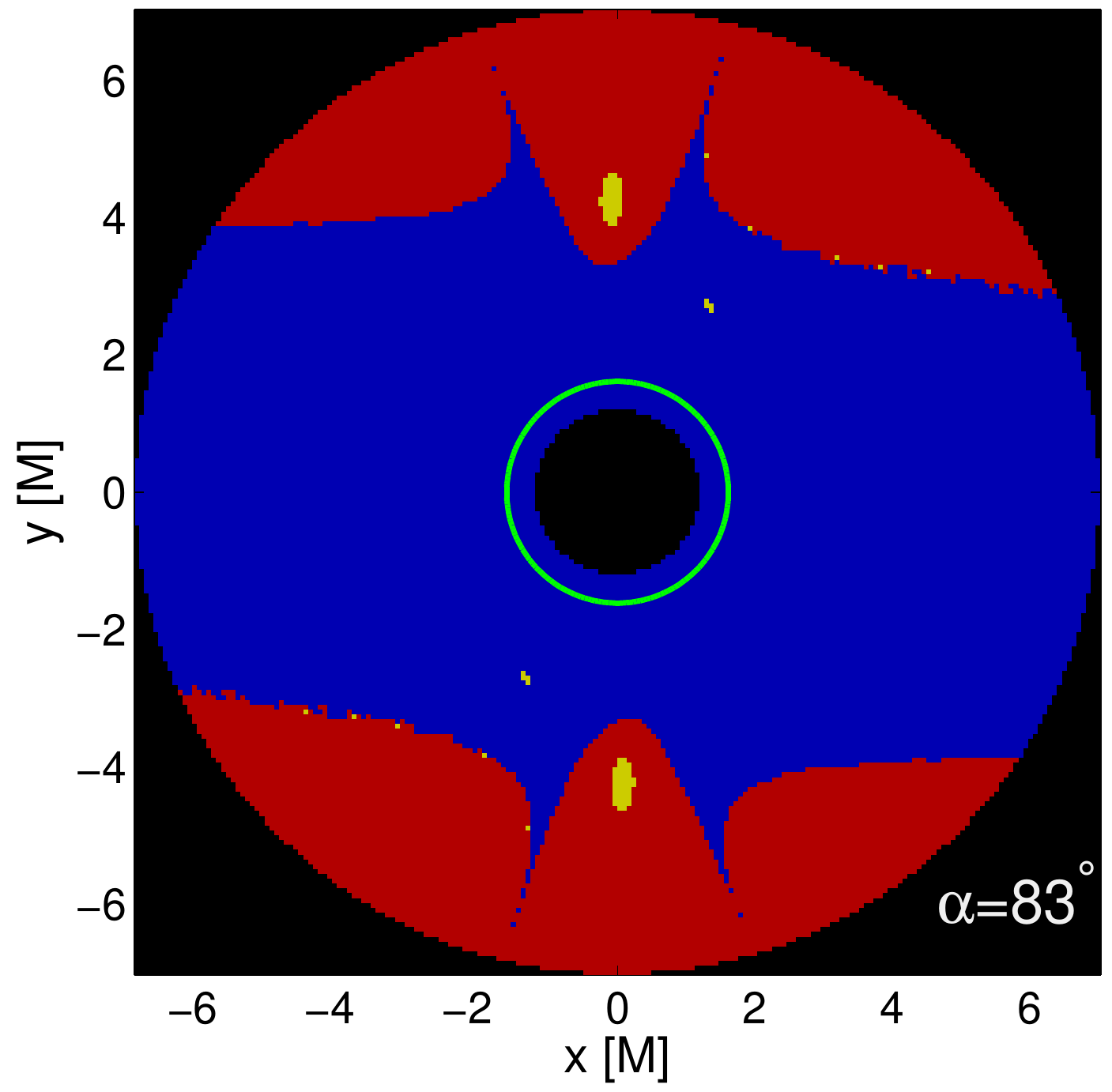}~~
\includegraphics[scale=.39]{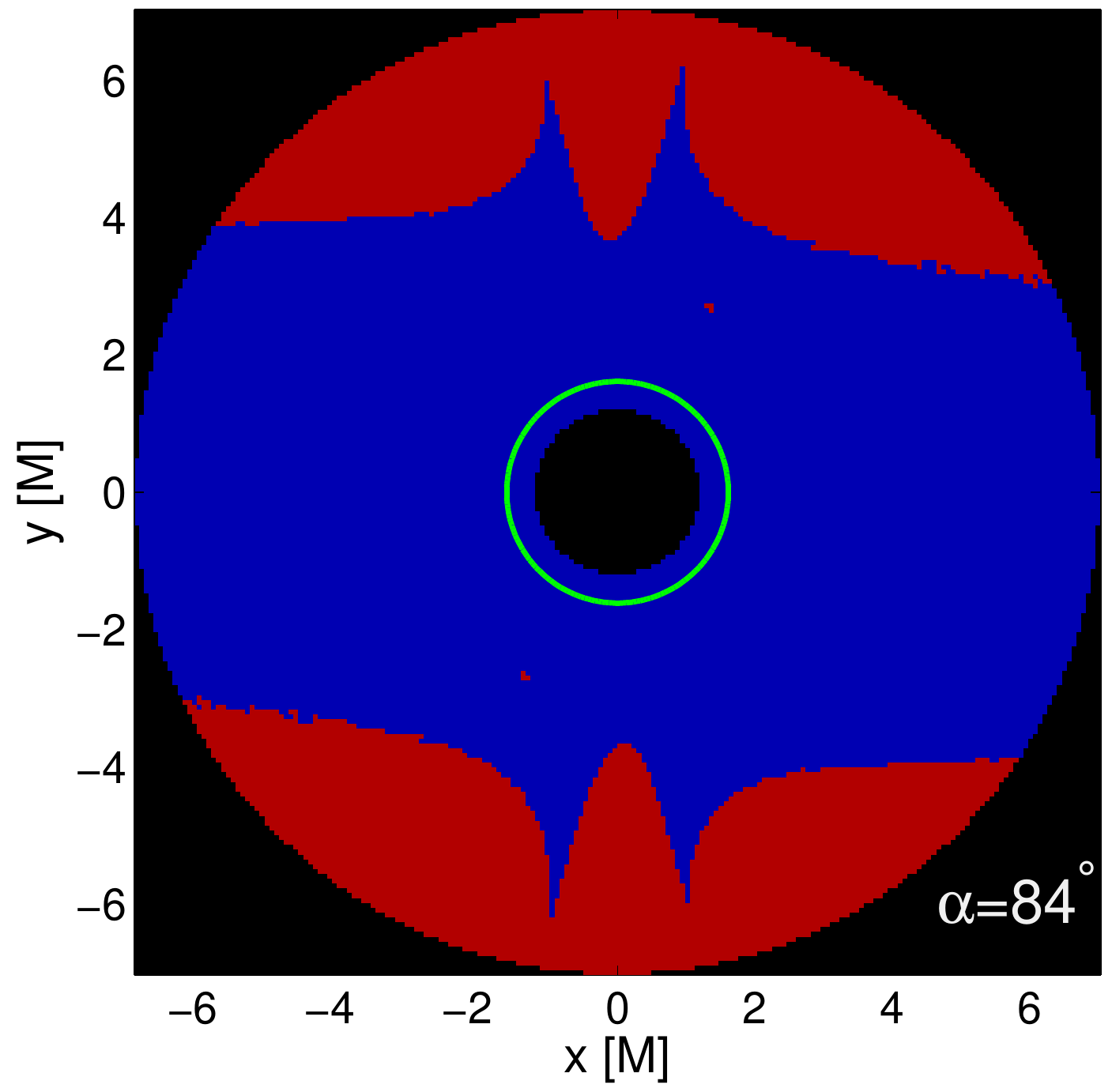}
\caption{Evolution and decline of the escape zones in the highly inclined magnetosphere. Color-coding and parameter choice are the same as in Figures~\ref{escape_primary_zone} and \ref{escape_23_zone}.}
\label{escape_decline}
\end{figure*}

For the case of equal values of the asymptotic field components ($B_z=B_x$, i.e., $\alpha=45\degree$), the primary escape zone becomes substantially deformed. If the inclination further increases, secondary and tertiary escape zones emerge and rise in size (Figure~\ref{escape_23_zone}). The maximum extent of the escape zones is reached for $\alpha\approx60\degree$, and with higher inclinations, the zones merge and gradually decline (top panels of Figure~\ref{escape_decline}). For $\alpha\gtrsim 80\degree$, the escaping orbits become rare until they vanish completely for $\alpha=84 \degree$ (bottom panels of Figure~\ref{escape_decline}).  

We may conclude that the inclination of the  magnetic field strongly supports the formation of the escape zones of charged particles. Analyzing the case for which there are no escaping orbits in the aligned setup, we have seen that even a very small inclination ($\alpha=1\degree$) is sufficient to induce the escaping trajectories; higher inclinations lead to the formation of large escape zones. However, as predicted by the analysis of the effective potential in Sec.~\ref{ionization}, the role of the aligned component $B_z$ is crucial, and the escape zones vanish as the inclination approaches $\alpha=\pi/2$.

\subsection{Acceleration and Terminal Velocity of Escaping Particles}
\label{acceleration}
In the previous section, we discussed the formation and evolution of equatorial escape zones with respect to the inclination angle $\alpha$ for a fixed value of magnetization $qB$ and spin $a$. In order to investigate the final velocity of escaping particles of four-velocity $u^{\mu}$ and to seek the most accelerated ones, we determine the linear velocity $v^{(i)}$ with respect to the locally nonrotating frame \citep{bardeen72} with the tetrad basis $e^{(i)}_{\mu}$ as follows:
\begin{equation}
\label{linspeed}
 v^{(i)}=\frac{u^{(i)}}{u^{(t)}}=\frac{e^{(i)}_{\mu}u^{\mu}}{e^{(t)}_{\mu}u^{\mu}}.
\end{equation}
We use these velocity components to express the Lorentz factor $\gamma=(1-v^2)^{-1/2}$, where $v=\sqrt{[v^{(r)}]^2+[v^{(\theta)}]^2+[v^{(\varphi)}]^2}$.

\begin{figure*}[ht]
\center
\includegraphics[scale=.48, trim={20mm 0 21mm 0},clip]{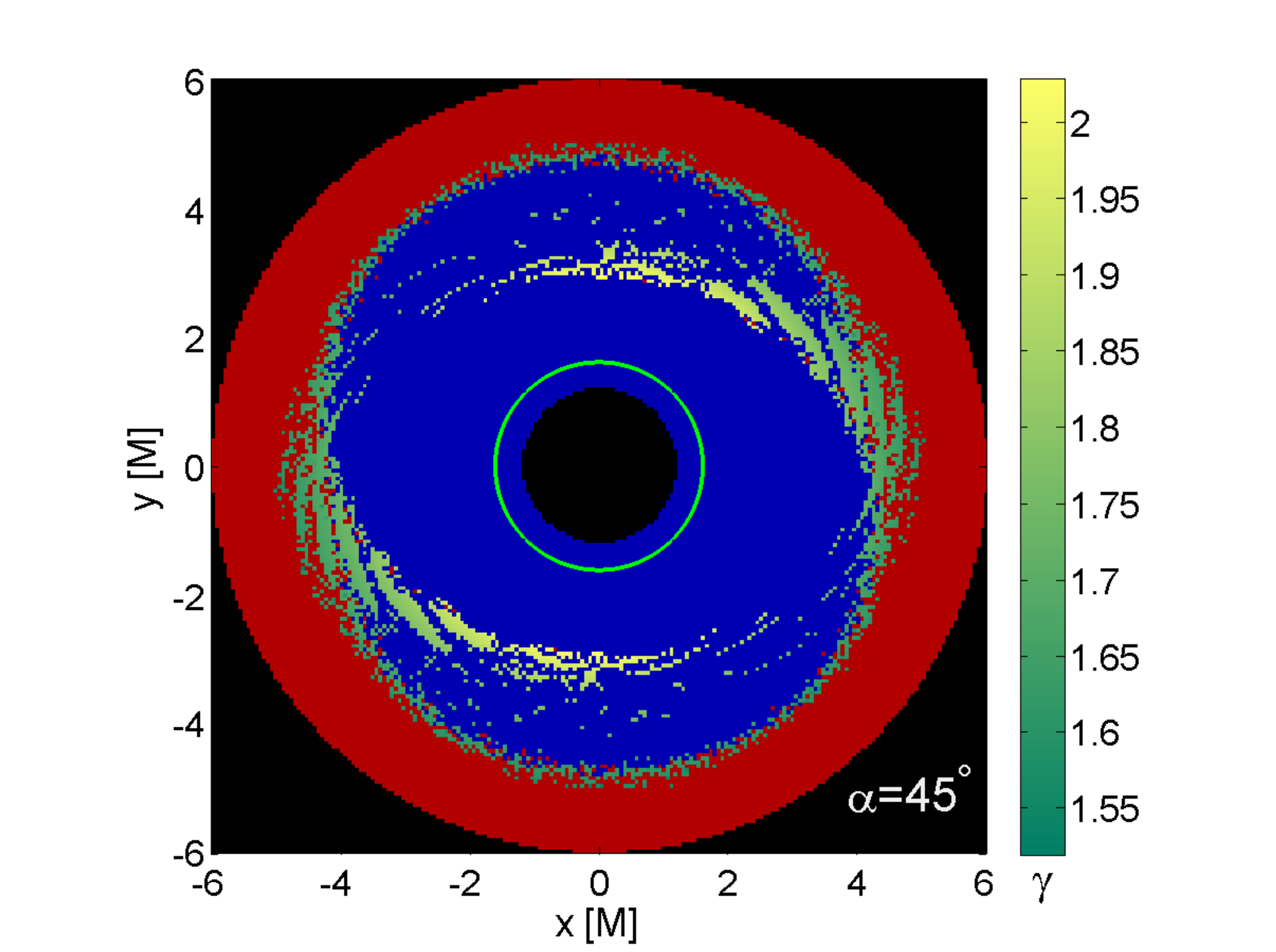}
\includegraphics[scale=.48, trim={15mm 0 21mm 0},clip]{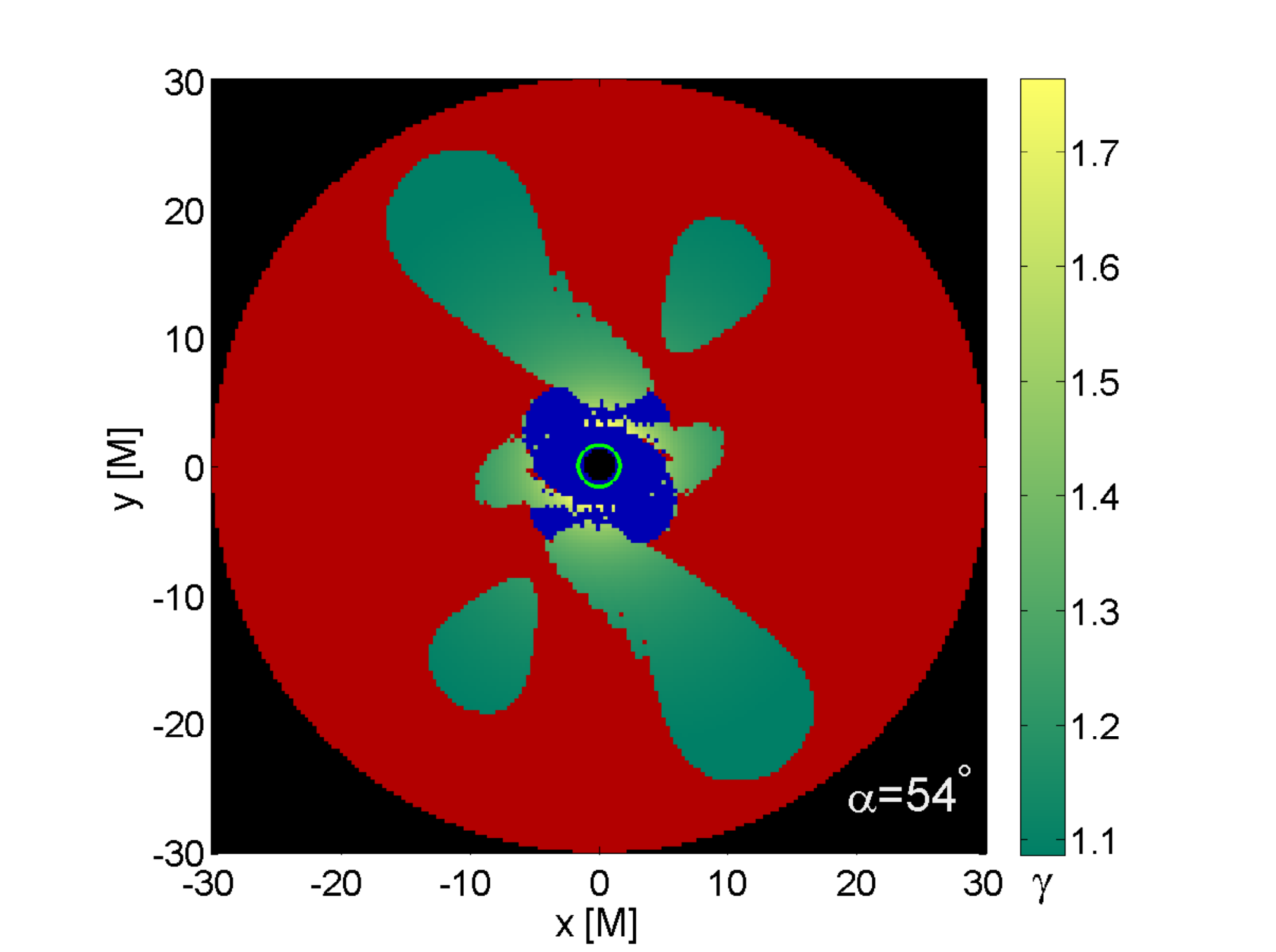}
\caption{Final Lorentz factor $\gamma$ of particles escaping from the equatorial plane, shown with the corresponding color scale. The parameter choice is the same as in Figures~\ref{escape_primary_zone}--\ref{escape_decline} ($qB=-5$, $a=0.98$). The field is inclined in the positive $x$-direction.}
\label{escape_acc}
\end{figure*}

The final value of $\gamma$ of trajectories in the escape zone is computed for the two examples of highly inclined magnetospheres analyzed in the previous section. We use the color scale to show the distribution of values of final $\gamma$ within the escape zones for $\alpha=45\degree$ (left panel of Figure~\ref{escape_acc}) and $\alpha=54\degree$ (right panel). As predicted by the analysis of effective potential in Sec.~\ref{ionization}, the maximal $\gamma$ is found for the trajectories with minimal $r_0$. We observe that minimal $r_0$ is achieved by particles with $\varphi_0 \approx \pi/2$, where the angle $\varphi$ is measured counter-clockwise from the positive direction of the $x$-axis (which is the direction of the field inclination). Comparing both panels of Figure~\ref{escape_acc} we also confirm that with a higher inclination of the field (of fixed magnitude $qB$), we obtain lower values of $\gamma$. While the perpendicular component $B_x$ is crucial as a perturbation that also allows the escape in cases where it was prohibited in the parallel field and supports the formation of extended escape zones, the acceleration of escaping particles is actually controlled by the parallel component $B_z$ (see Equation.~(\ref{asym})).

The maximal Lorentz factor $\gamma_{\rm max}$ that can be achieved by escaping particles in the aligned field was investigated in Paper~I. In particular, we found that the value of $\gamma_{\rm max}$ saturates at $\gamma_{\rm max}\approx6$, which is attained with $|qB|\gtrapprox 100$ for $a\lessapprox 0.1$, while the maximally spinning black hole accelerates the escaping matter only up to $\gamma_{\rm max}\approx2.5$ for $qB\approx-4.5$. However, as demonstrated in the previous section, the inclination of the field also supports the formation of escape zones around rapidly spinning black holes. On the other hand, the analysis of the effective potential (\ref{eff_pot}) suggests that $\gamma$ increases with $B_z$ and with fixed $B$; it thus decreases as $\alpha$ grows. Maximally accelerated particles are therefore expected from a slightly inclined magnetosphere, where the inclination $\alpha$ is sufficient to support the escape zone; however, the parallel component $B_z$, which is the  actual source of energy for the acceleration, remains dominant. 

\begin{figure}[ht]
\center
\includegraphics[scale=.5, trim={8mm 0 0mm 0}]{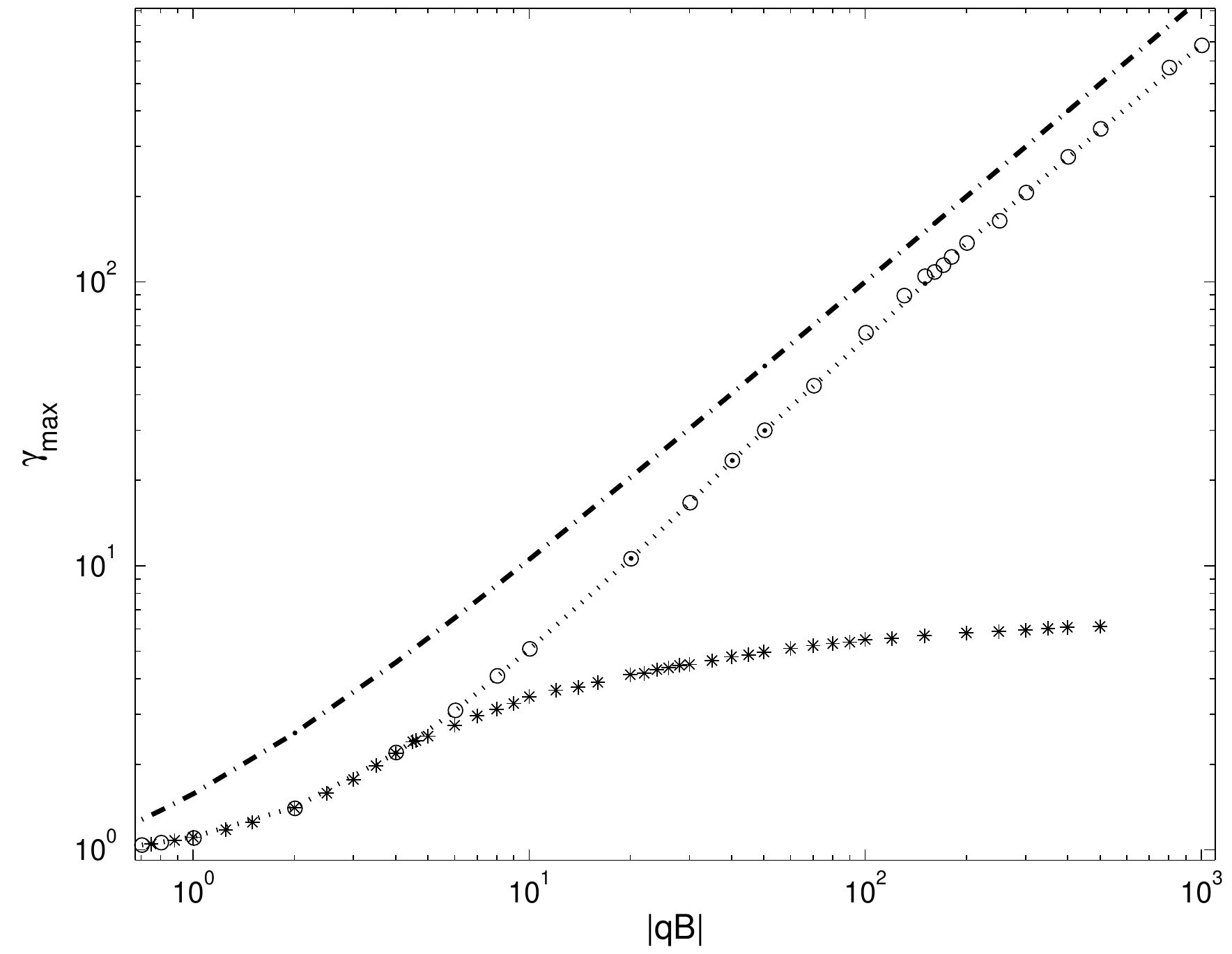}
\caption{Maximal value of the final Lorentz factor of escaping particles as a function of magnetization parameter $|qB|$. Circles denote the values obtained numerically for trajectories in the inclined magnetosphere ($B_{x}/B_{z}=0.1$, i.e., $\alpha\approx 6^{\degree}$ with $\varphi_0 = \pi/2$), while asterisks show the aligned case ($B_x =0$) analyzed in Paper~I. The dotted line shows the expected value predicted by Equation~(\ref{lorentz}) for the numerically obtained value of $r^{\rm min}_0$, and the dashed--dotted line is the theoretical maximum for $r_0 = r_+ =1$.}
\label{escape_gamma}
\end{figure}

For the above reasons, we restrict our search for maximally accelerated escaping particles to the small inclination of the field (we set $B_x/B_z=0.1$, i.e., $\alpha\approx6\degree$), extremal spin $a=1$, and $\varphi_0=\pi/2$ in order to support the escaping trajectories of highly accelerated particles from the small radii. The expected value of the final Lorentz factor $\gamma$ may be derived from Equation~(\ref{asym}) as

\begin{equation}
\label{lorentz}
\gamma(a,r_0,|qB_z|)=E_{\rm Kep}(a,r_0)+\frac{|qB_z|a}{r_0}.
\end{equation}

However, the values of $r_0$ for which the escaping orbits are actually realized are not known a priori, and we have to numerically investigate the trajectories in the relevant range of initial radii to localize the inner edge of the escape zone, i.e., to identify the escaping trajectory with  minimal value $r^{\rm min}_0$ corresponding to $\gamma_{\rm max}$. Numerical analysis confirms that $r^{\rm min}_0$ generally decreases as $|qB|$ grows. In particular, for the given parameters ($B_x/B_z=0.1$, $\varphi_0=\pi/2$), we observe that while $|qB|\approx1$ leads to $r^{\rm min}_0\approx5$, it gradually decreases to $r^{\rm min}_0\approx 2$ for $|qB|\approx10$ and eventually stabilizes at $r^{\rm min}_0\approx1.5$ for $|qB|\gtrapprox100$. In Figure~\ref{escape_gamma} we show the corresponding values of $\gamma_{\rm max}$  as a function of $|qB|$ and compare them with the analogous data for the aligned magnetic field from Paper~I. They both coincide for small values of $|qB|\lessapprox 4.5$; however, then the curves split as the stronger aligned field excludes the escape from the vicinity of rapidly spinning black holes and the acceleration saturates at $\gamma_{\rm max}\approx 6$. The inclined field seems to allow the escape from the maximally spinning hole for any value of $|qB|$, and $\gamma_{\rm max}$ grows steadily with the linear trend suggested by Equation~(\ref{lorentz}). The actual values of $\gamma_{\rm max}$ are compared with the values predicted for a given numerically obtained $r_0$, and they agree very well (dotted line in Figure~\ref{escape_gamma}). The dashed--dotted line shows the theoretical maximum of $\gamma_{\rm max}$ obtained by substituting  $a=1$ and $r_0=r_+=1$ in Equation~(\ref{lorentz}).

Breaking the axial symmetry appears to have profound consequences regarding the acceleration of escaping particles. While the axisymmetric setup allows the acceleration up to $\gamma \lessapprox 6$ (and even more strictly limited for rapidly rotating black holes), here, with the oblique magnetosphere of sufficient $|qB|$, we may obtain ultrarelativistic particles with $\gamma\gg1$. In particular, we numerically confirmed the value of $\gamma=680$ for $|qB|=10^3$. Despite numerical difficulties in tracing it in strong fields, $\gamma$ is supposed to further grow with increasing $|qB|$ and does not seem to have any fixed limit, since radiation losses \citep{tursunov18} are not considered in the present analysis. The small inclination of the field thus proves sufficient to perturb the dynamics of stable orbits, and, unlike the aligned field, it admits the acceleration of particles to ultrarelativistic velocities. The efficiency of the process grows with the spin, reaching its maximum for extremal black holes; however, large $\gamma$ may also be obtained for moderately rotating black holes.          

\begin{figure*}[ht]
\center
 \includegraphics[scale=.42]{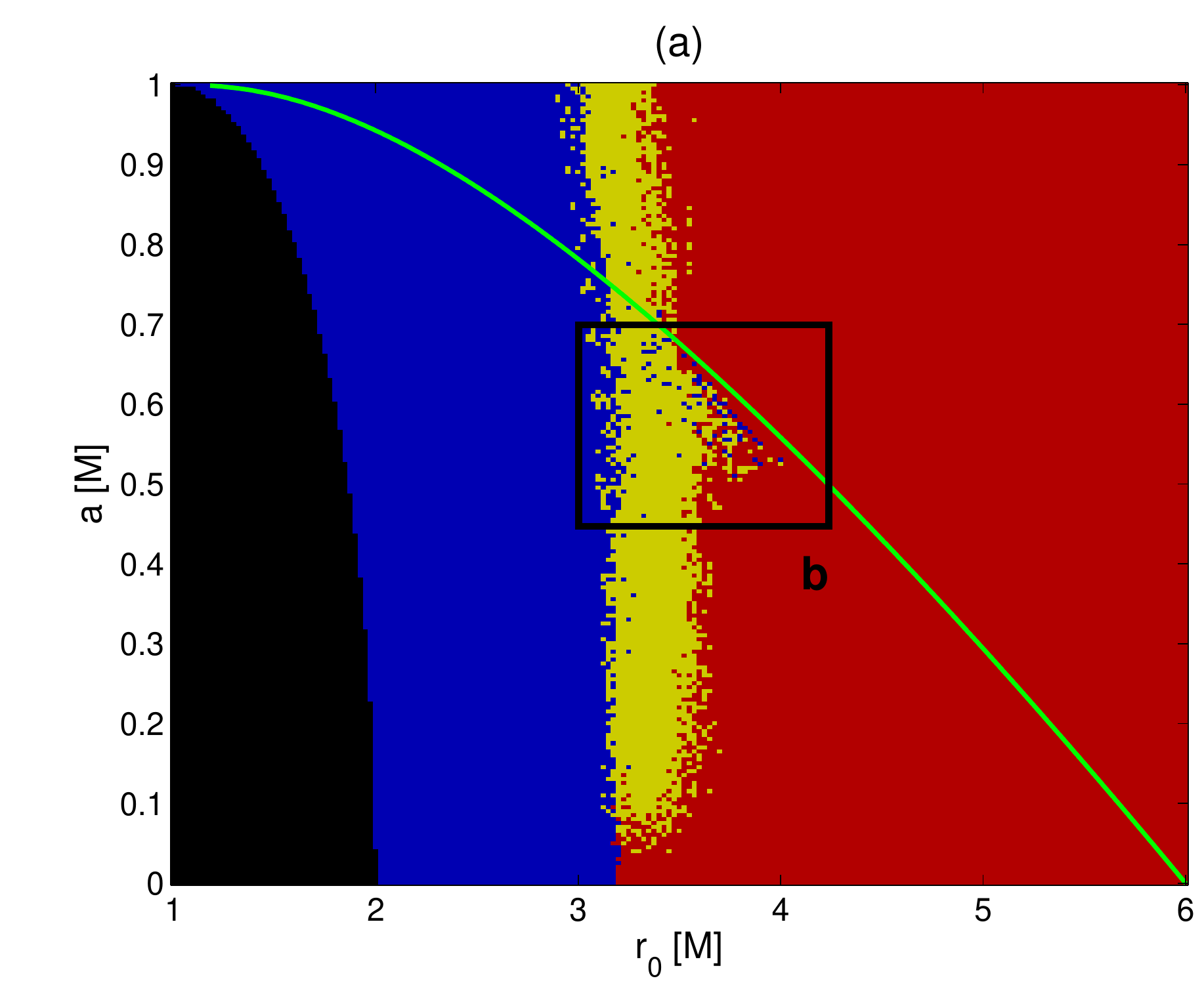}
 \includegraphics[scale=.42]{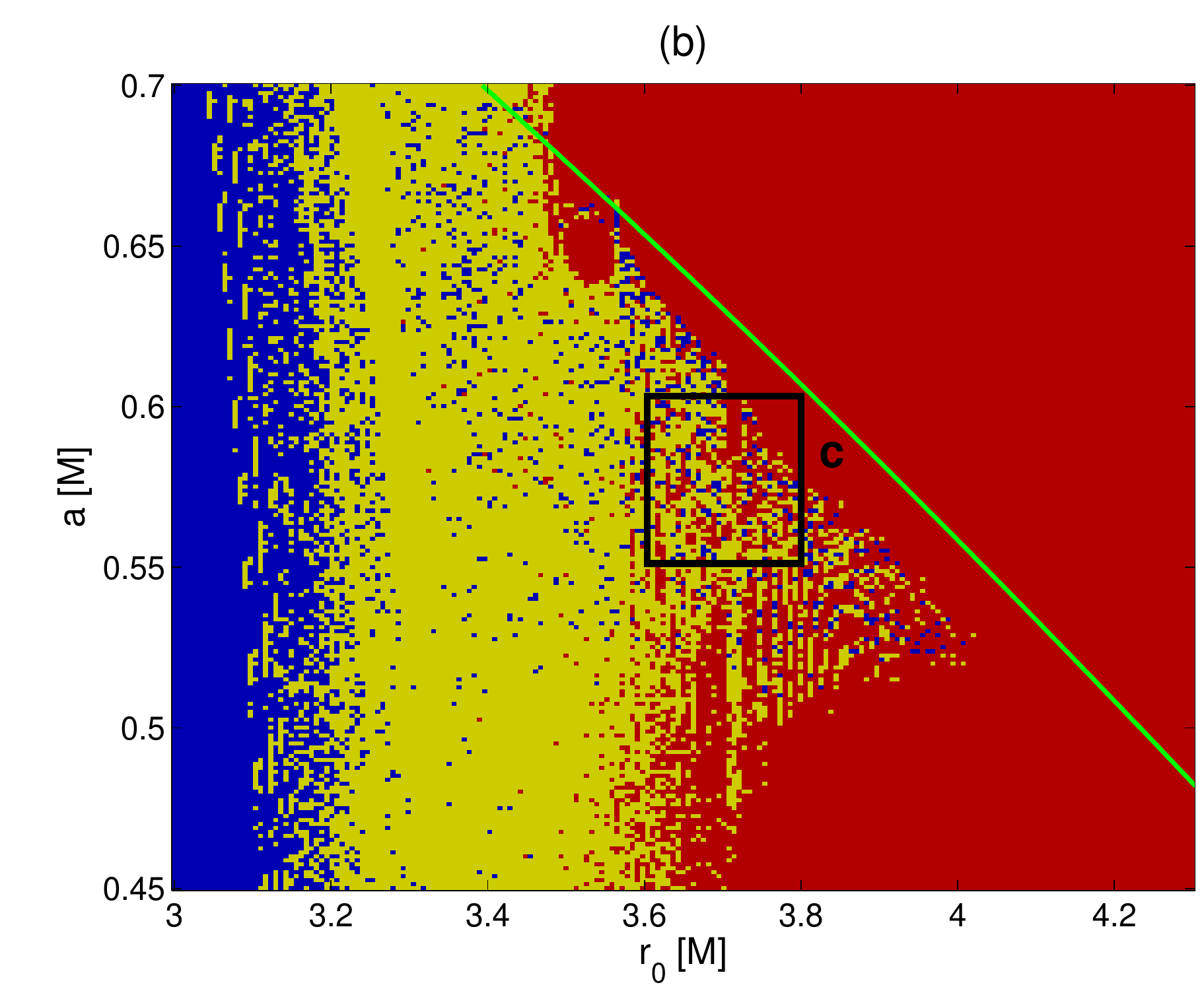}\\
\includegraphics[scale=.42]{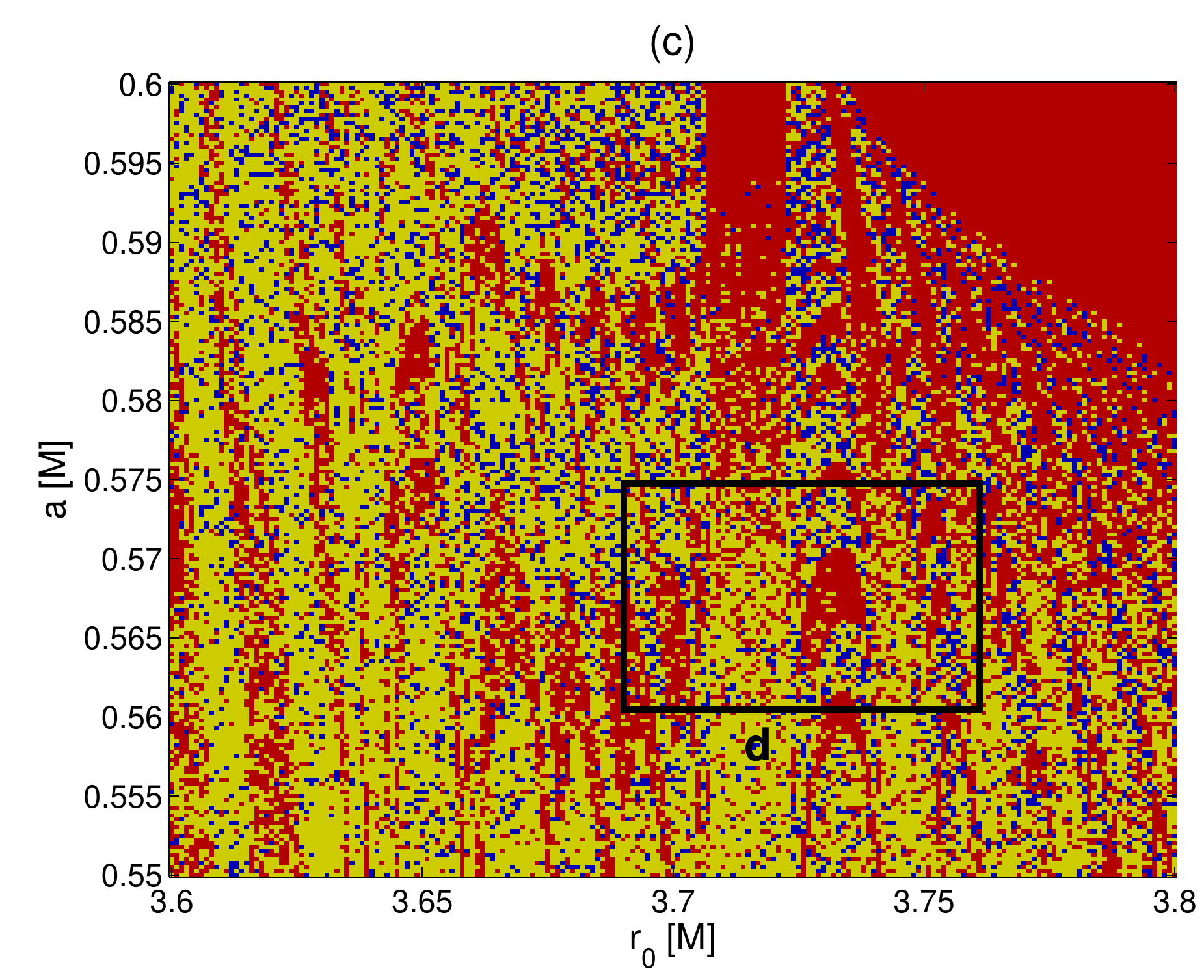}
 ~\includegraphics[scale=.42]{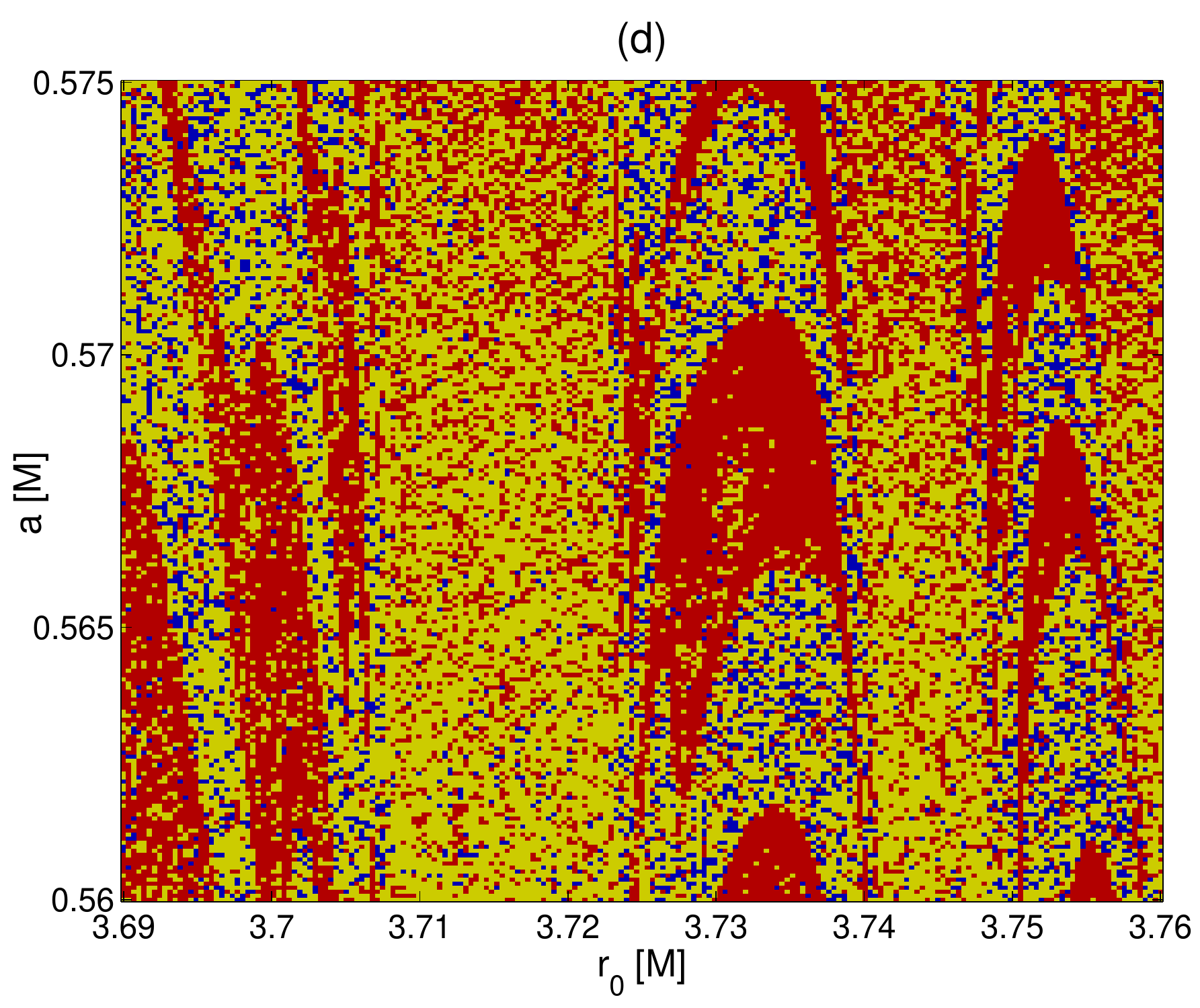}
\caption{The structure of the escape zone ($qB=-4.1$, $\alpha=14\degree$, and $\varphi_0=0$) in the relevant range of the spin parameter $a$ and initial radius $r_0$ is explored. Going from the top left to the bottom right panel, the portions of the plots (marked by black rectangles) are  magnified progressively in the subsequent panels, revealing the complex structure. The green line represents the ISCO, and black color shows the horizon of the black hole. Color-coding of trajectories is the same as in Figures~\ref{escape_primary_zone}--\ref{escape_decline}.}
\label{escape_detail1}
\end{figure*}

\subsection{Transient Chaos and Fractal Geometry of Escape Zones}
\label{chaos}
The escape zones in the axisymmetric setup studied in Paper~I revealed a complicated inner structure; in particular, we found regions with self-similar patterns at different magnification (Figure~7 in Paper~I). The escape zone was typically composed of escaping orbits intermixed with stable orbits in a manner characteristic of objects with fractal geometry described by a noninteger dimension. This observation, together with other indications, suggests that deterministic chaos plays an important role in the escape zone, and transient chaos is a key ingredient in the dynamics of escaping particles. We expect that, compared to the previous axisymmetric case, the amount of chaos increases when the symmetry breaks, since we have previously performed the analysis of bound orbits in this system \citep{kopacek14} clearly showing the growth of chaos due to increasing inclination.

Nevertheless, chaotic episodes of unbound escaping orbits are rather short, as the particle quickly escapes far enough from the vicinity of the black hole to the region where the spacetime is almost flat and magnetic field almost uniform; thus, the nonintegrable perturbation that induces chaotic behavior diminishes there. For this reason, it becomes problematic to quantify these trajectories by the standard chaotic indicators given by the Lyapunov characteristic exponents \citep[applied for bound orbits in][]{kopacek14}, since these typically require a very long time to converge \citep{skokos10}. Also, the visualization of the trajectories in the Poincar\'{e} surfaces of section is not effective here, since the unbound motions of escaping trajectories do not provide enough intersection points. Moreover, the nonaxisymmetric system has three degrees of freedom, which further reduces the applicability of this method. Therefore, we need to employ alternative tools, e.g., the technique of recurrence analysis \citep{marwan07}, which is able to indicate the chaotic behavior on a substantially shorter time-scale regardless of the dimensionality of the system.

In Figure~\ref{escape_detail1} we study the details of a particular escape zone with $qB=-4.1$, inclination $\alpha=14\degree$, and initial azimuthal angle $\varphi_0=0$. The escape zone is plotted in the $r_0\times a$ plane (ionization radius vs. spin) with the same color-coding of trajectories as in Figures~\ref{escape_primary_zone}-\ref{escape_decline} (blue for plunging, yellow for escaping, and red for oscillating).\footnote{In Paper~I, we have denoted such visual representation of the system as a {\em basin-boundary plot} in the analogy of three different final states with the attractors of the dynamical system. However, since real attractors are only found in dissipative systems, and their basins of attraction are usually represented with the phase space variables, we do not use this term further for our plots (with the spin parameter $a$ as a variable) to avoid confusion. In the context of escaping trajectories in a conservative system, the term {\em escape-boundary plot} could be used instead \citep{tel06}.}  We observe the complex structure of the primary escape zone, which is characteristic of objects with the fractal geometry. In particular, we notice the regions where all three types of trajectories intermix, while in the axisymmetric case, the blue region of plunging orbits is always connected and has a well-defined boundary. The inclination of the field thus substantially increases the level of complexity of the primary escape zone, which indicates the presence of strong chaos.

\begin{figure*}[ht]
\center
 \includegraphics[scale=.4]{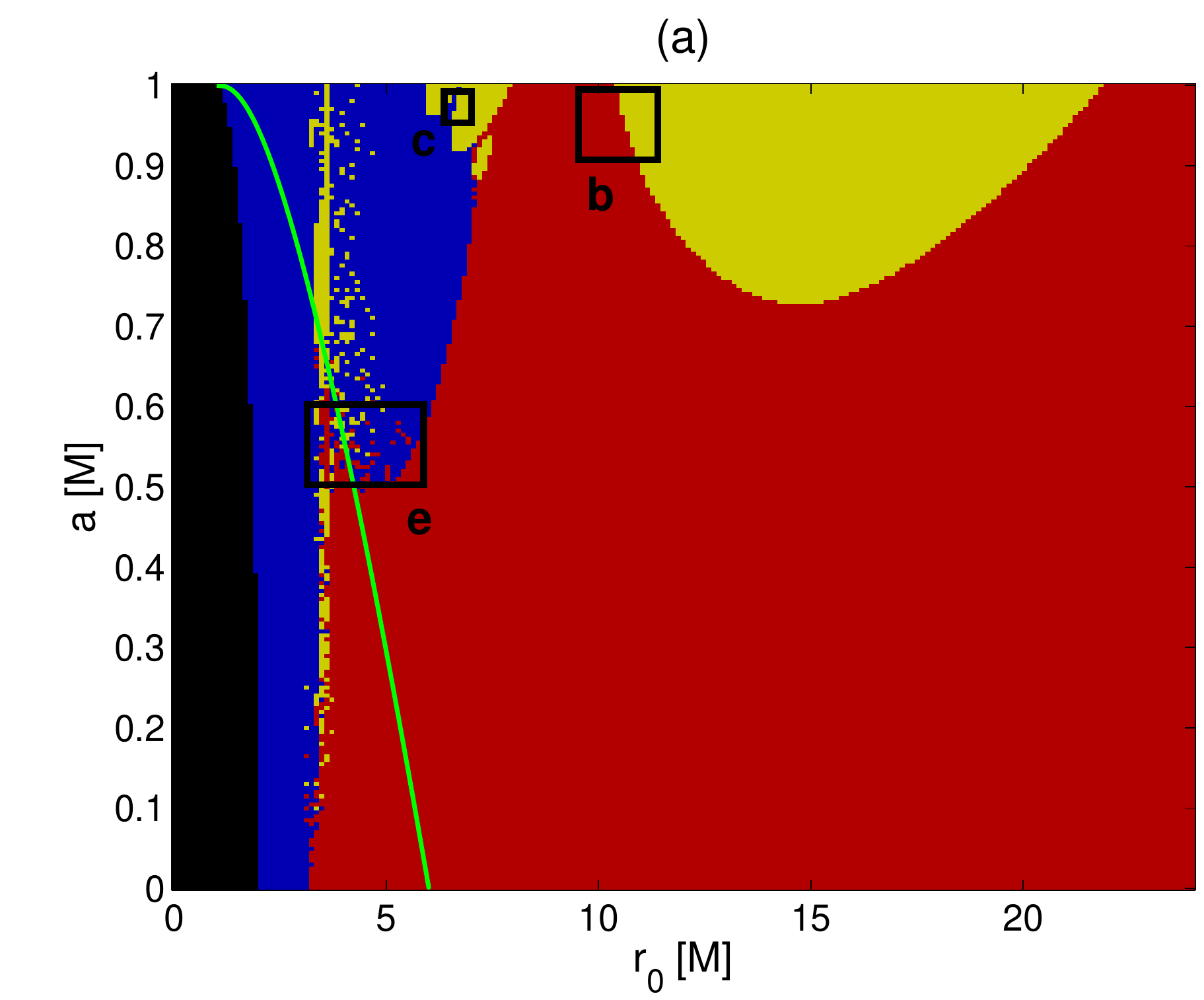}
~~\includegraphics[scale=.4]{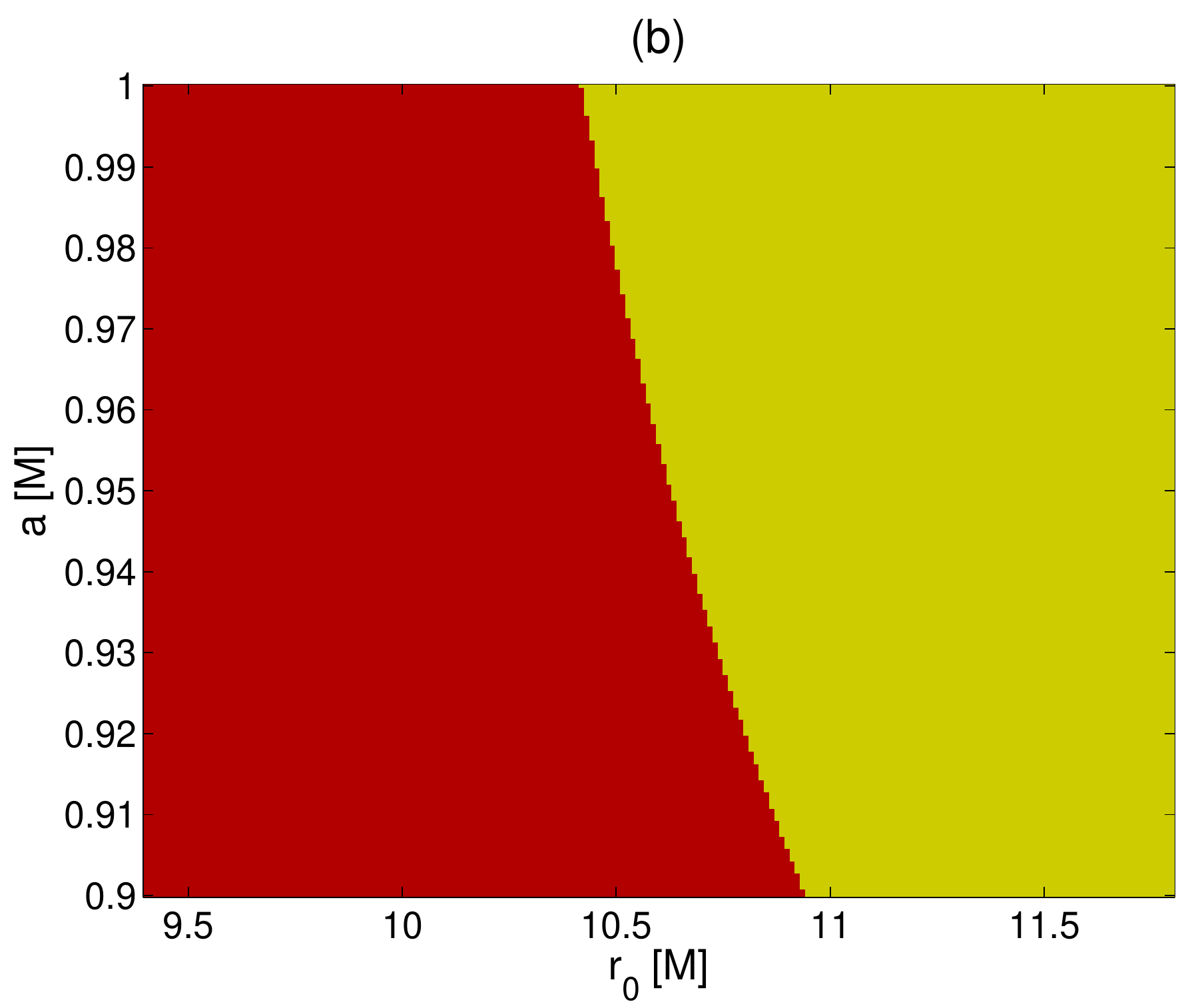}\\
~ \includegraphics[scale=.4]{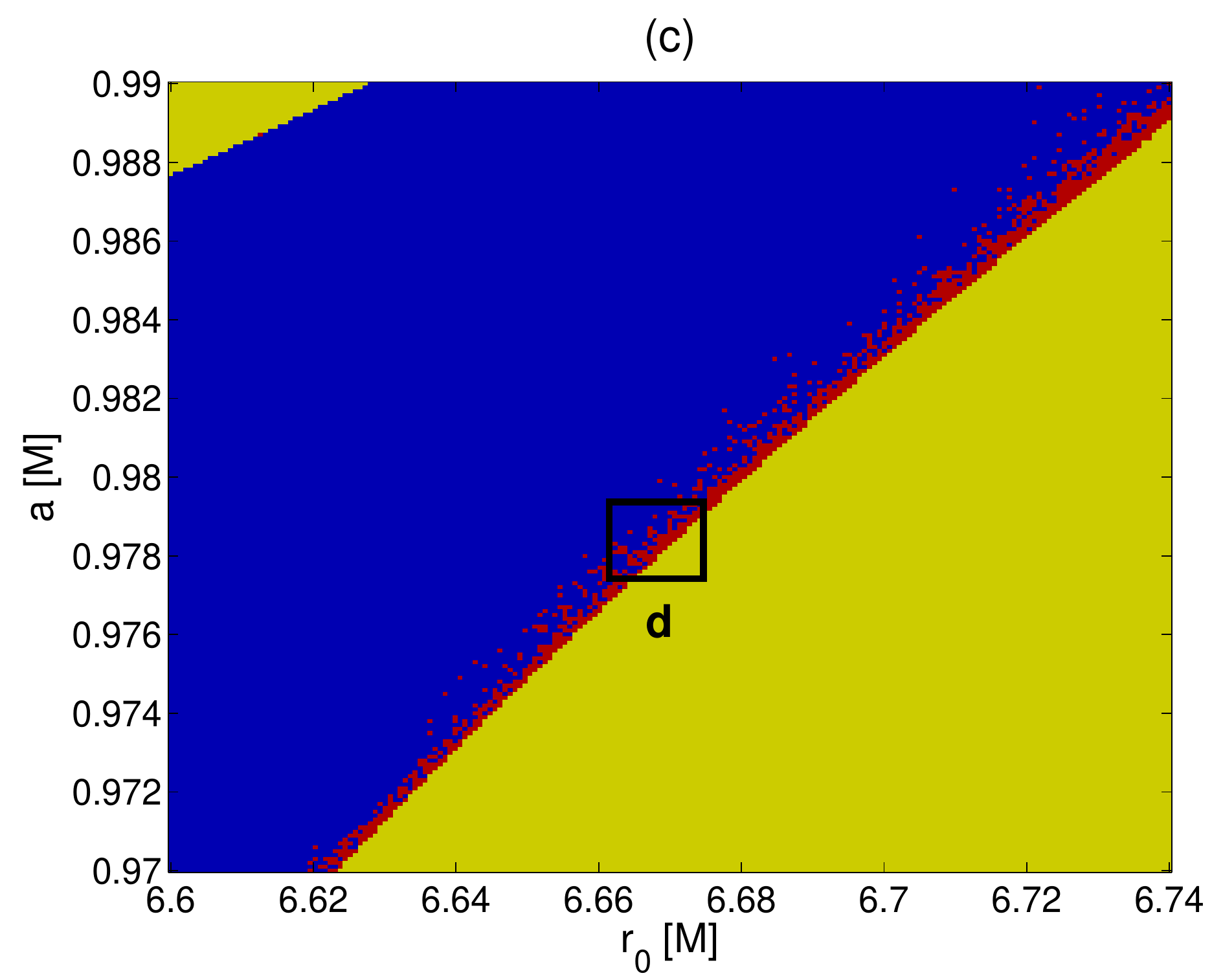}
 \includegraphics[scale=.4]{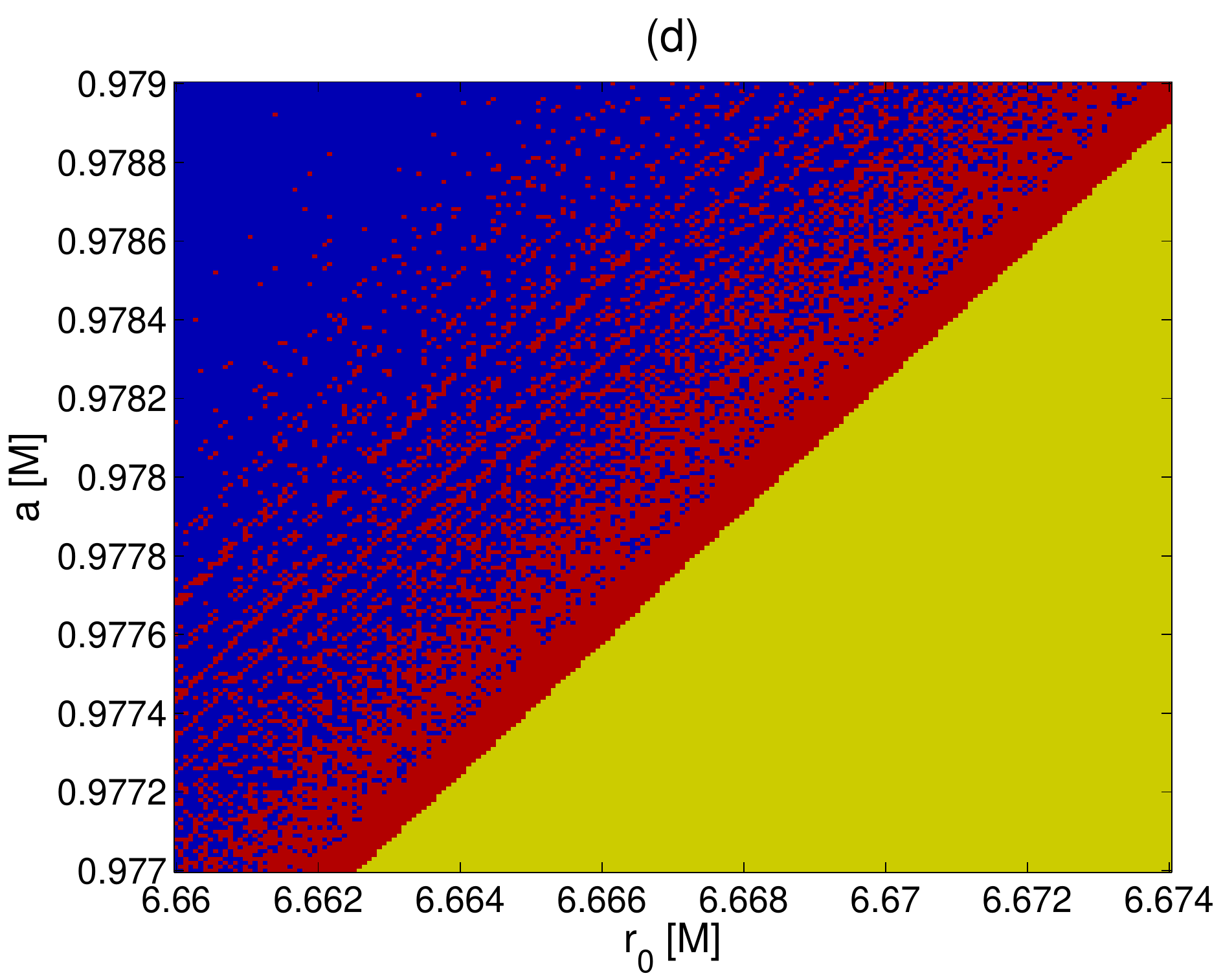}\\
\includegraphics[scale=.4]{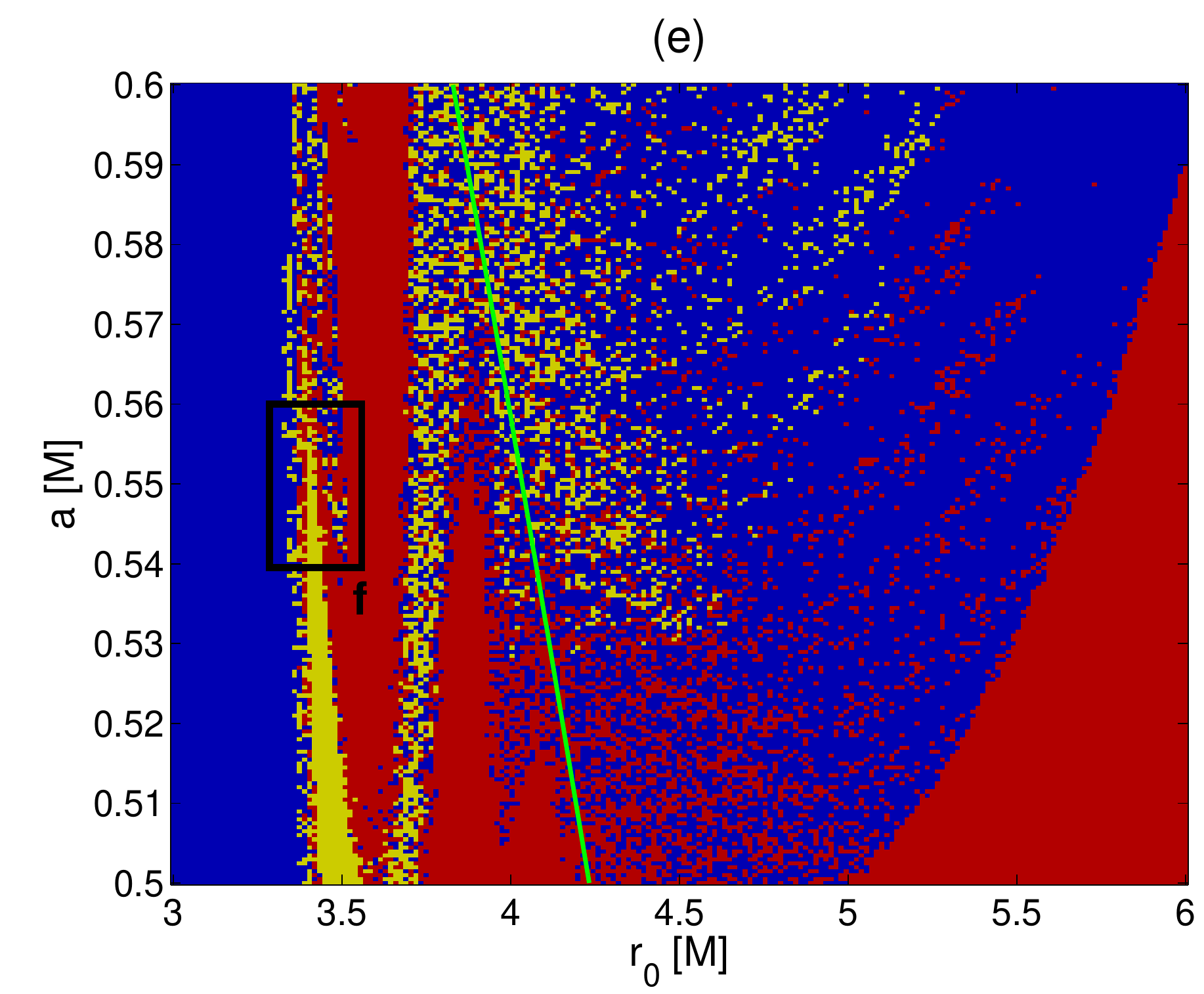}
 \includegraphics[scale=.4]{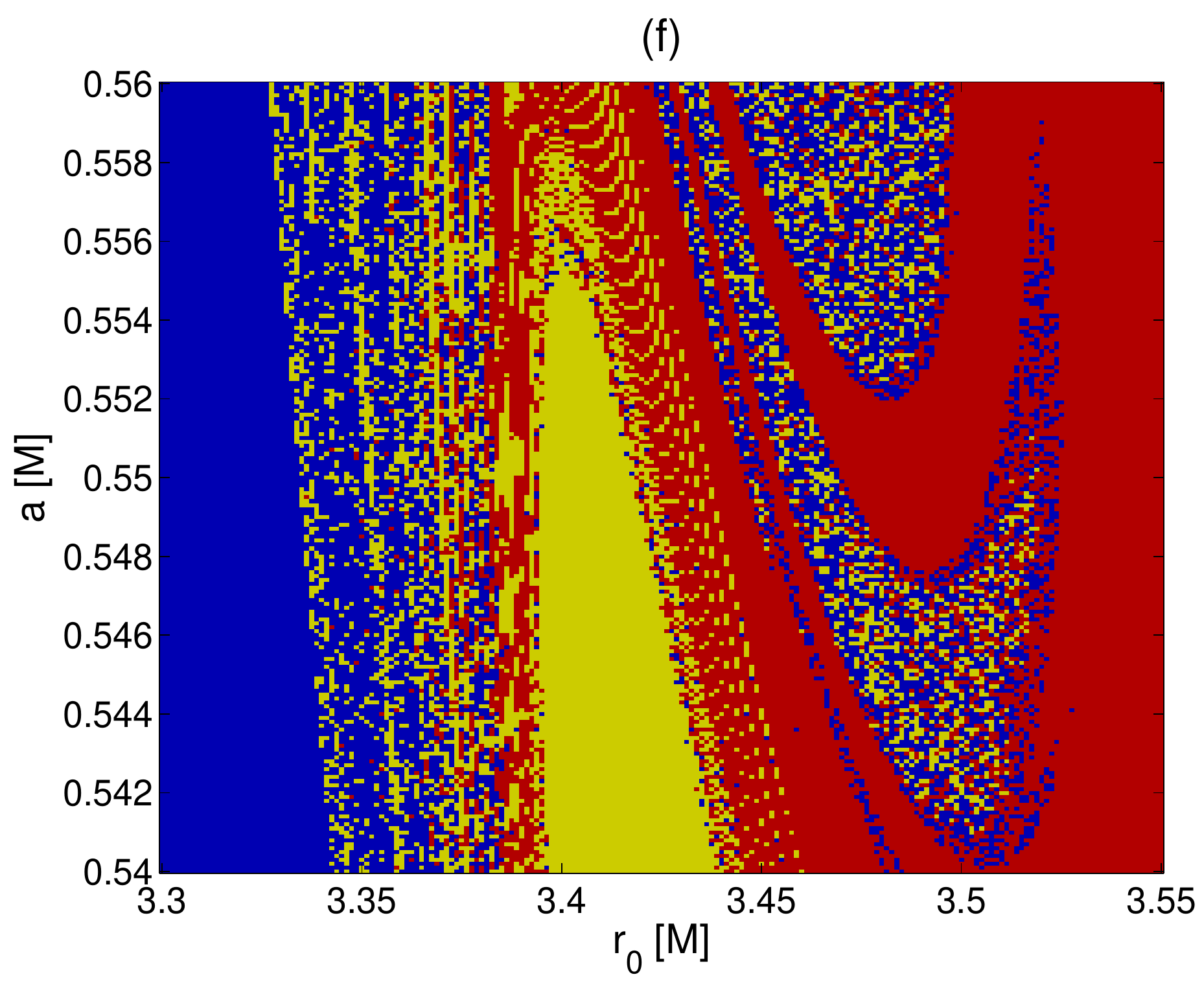}
\caption{Escape zones for the case $qB=-5$, $\alpha=54\degree$, and $\varphi_0=60\degree$. The tertiary escape zone (b) shows a well-defined edge, while the edge of the secondary escape zone (c, d) reveals a fuzzy layer with plunging and stable orbits intermixed. The primary escape zone (e, f) has a complex fractal structure with regions where all three types of trajectories intermix.}
\label{escape_detail2}
\end{figure*}

In Figure~\ref{escape_detail2}, we explore the structure of the escape zones for the highly inclined magnetosphere ($\alpha=54\degree$, $qB=-5$), where all the three classes of the escape zones are present (as already shown in  the last panel of Figure~\ref{escape_23_zone}). An appropriate choice of the initial azimuthal angle ($\varphi_0=60\degree$) allows one to depict them in a single $r_0\times a$ plot (panel (a) of Figure~\ref{escape_detail2}). The tertiary escape zone (panel (b)) reveals the well-defined boundary without traces of nonlinear effects. On the other hand,  the edge of the secondary escape zone (panels (c), (d)) shows a narrow fuzzy layer where plunging and stable orbits intermix, while the region of escaping orbits remains connected. Nevertheless, in the primary escape zone (panels (e), (f)), all three types of orbits intermix, and their domains are interwoven in a complex way with a fractal structure characteristic of chaotic systems. 

\subsection{Chaotic Indicators: Recurrence Analysis and Box-counting Dimension}
\label{rp}
Visual survey of the escape zones (Figures~\ref{escape_detail1} and \ref{escape_detail2}) suggests that chaos plays an important role mostly in the primary escape zone (the only zone encountered with small to moderate inclinations), while there is a minor indication of chaotic dynamics on the boundary of the secondary escape zone and no traces of chaos in the tertiary zone. In the following, we employ several quantitative indicators of chaos in order to further inspect this conjecture. Since we are dealing with transient chaos of escaping trajectories, standard tools like Lyapunov exponents or Poincar\'{e} sections become ineffective. Therefore, we employ the technique of recurrence analysis \citep{marwan07}, which allows one to detect chaotic behavior on a short time-scale. This method is based on the statistical analysis of the recurrences of the trajectory to the neighborhood of its previous states in the phase space. In particular, the binary-valued recurrence matrix $\mathbf{R}_{ij}$ is constructed as follows:

\begin{figure*}[ht]
\center
\includegraphics[scale=.34, trim={15mm 0 2mm 0},clip]{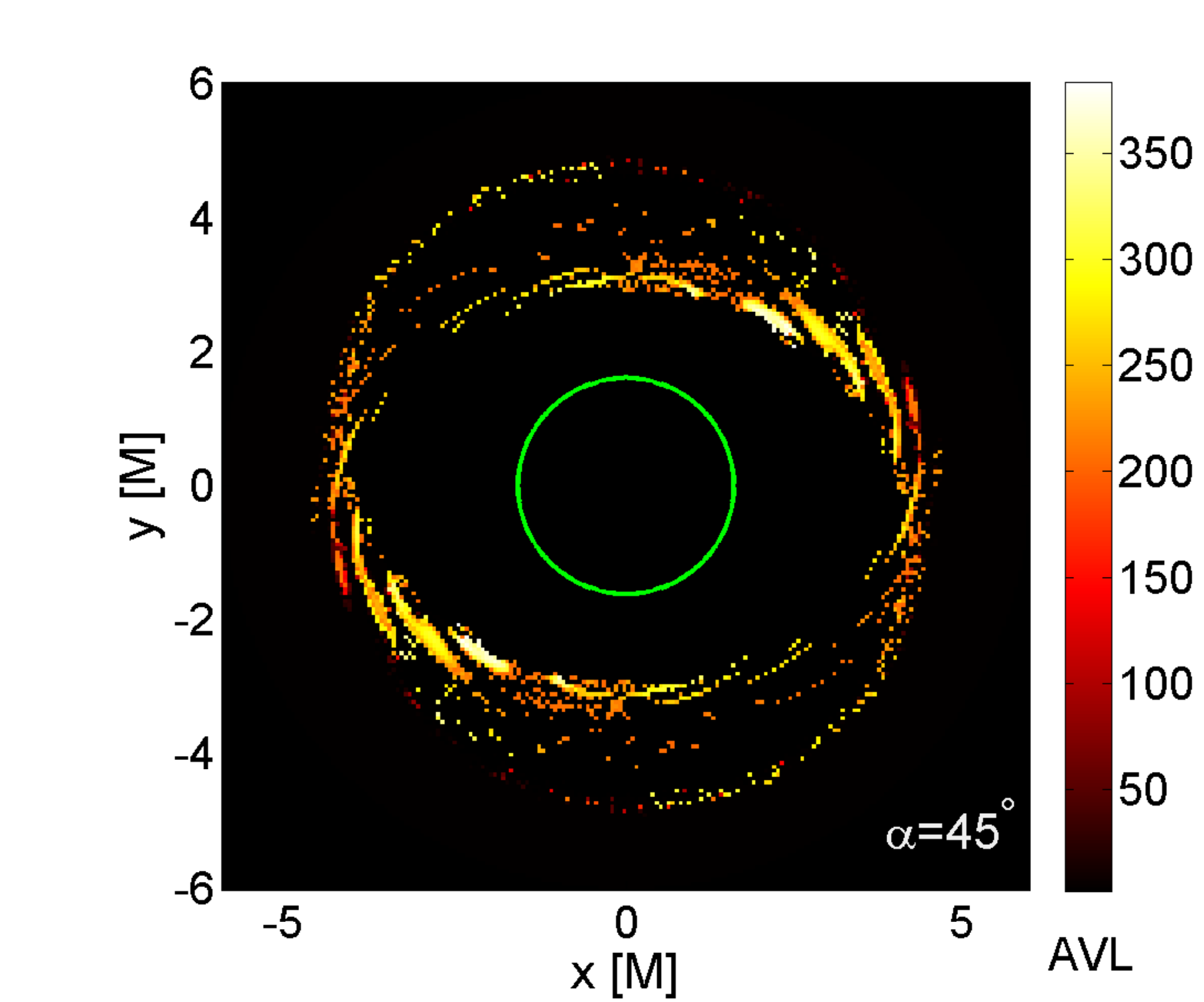}
\includegraphics[scale=.34, trim={15mm 0 2mm 0},clip]{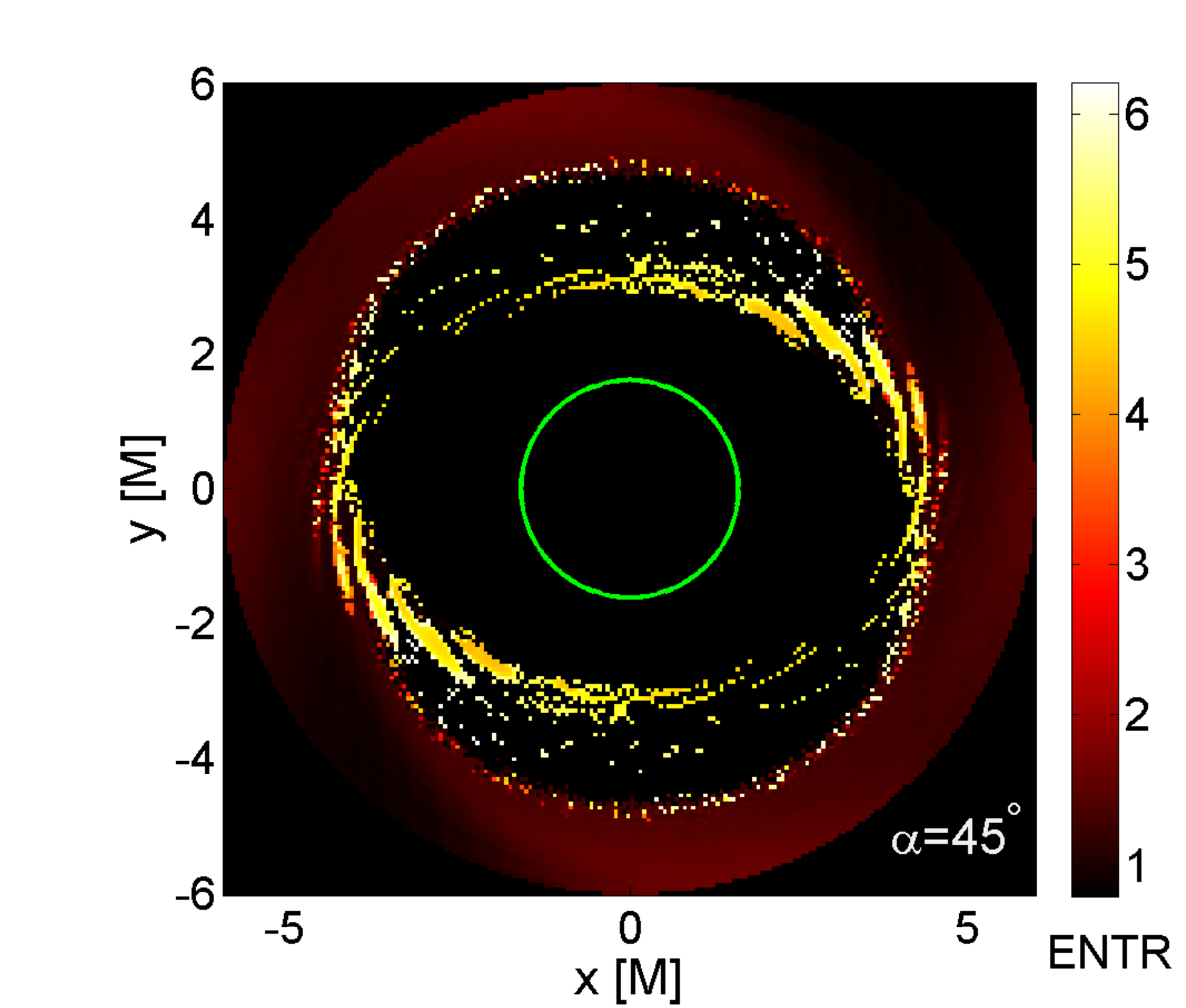}
\includegraphics[scale=.34, trim={12mm 0 2mm 0},clip]{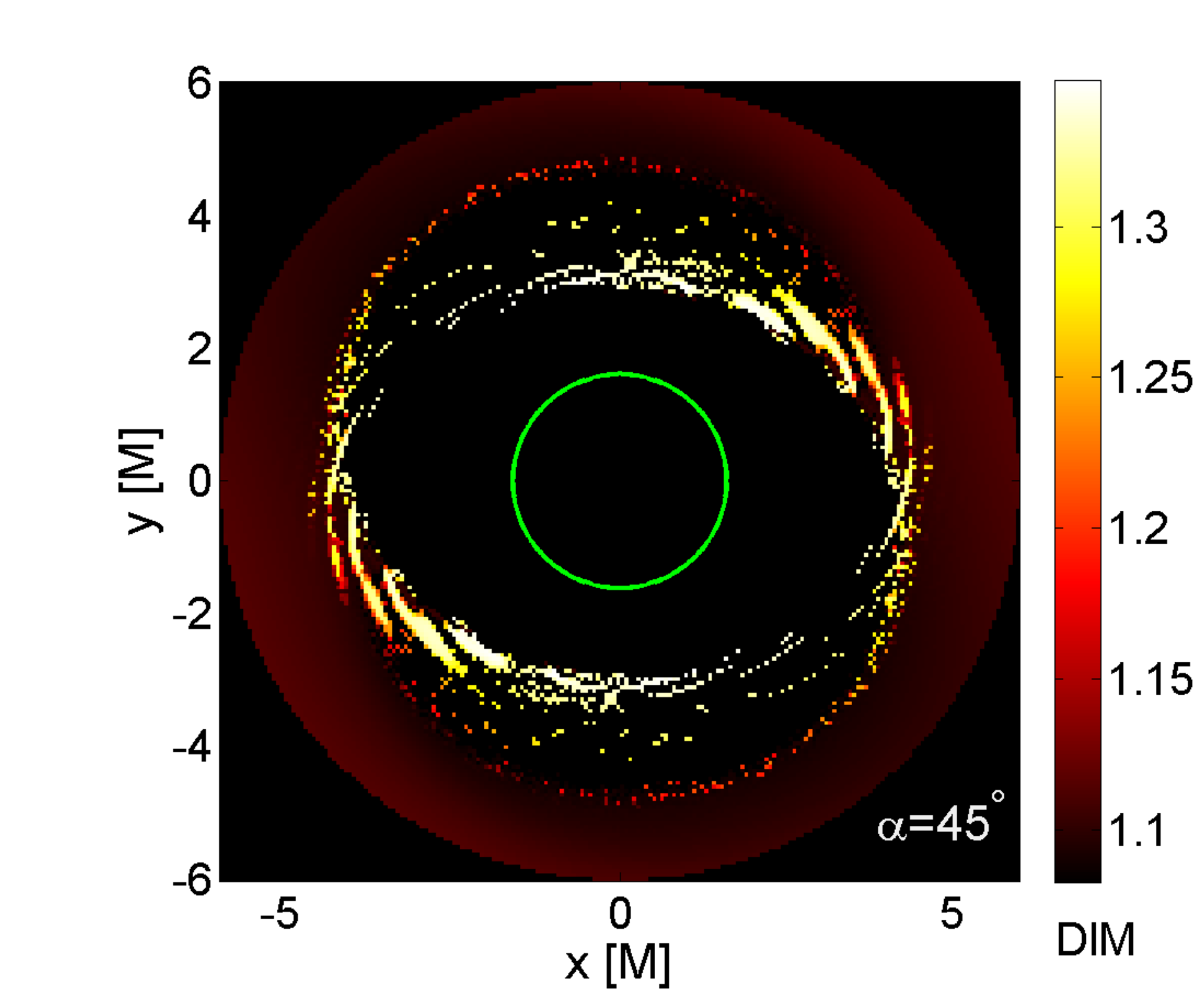}\\
\includegraphics[scale=.34, trim={10mm 0 2mm 0},clip]{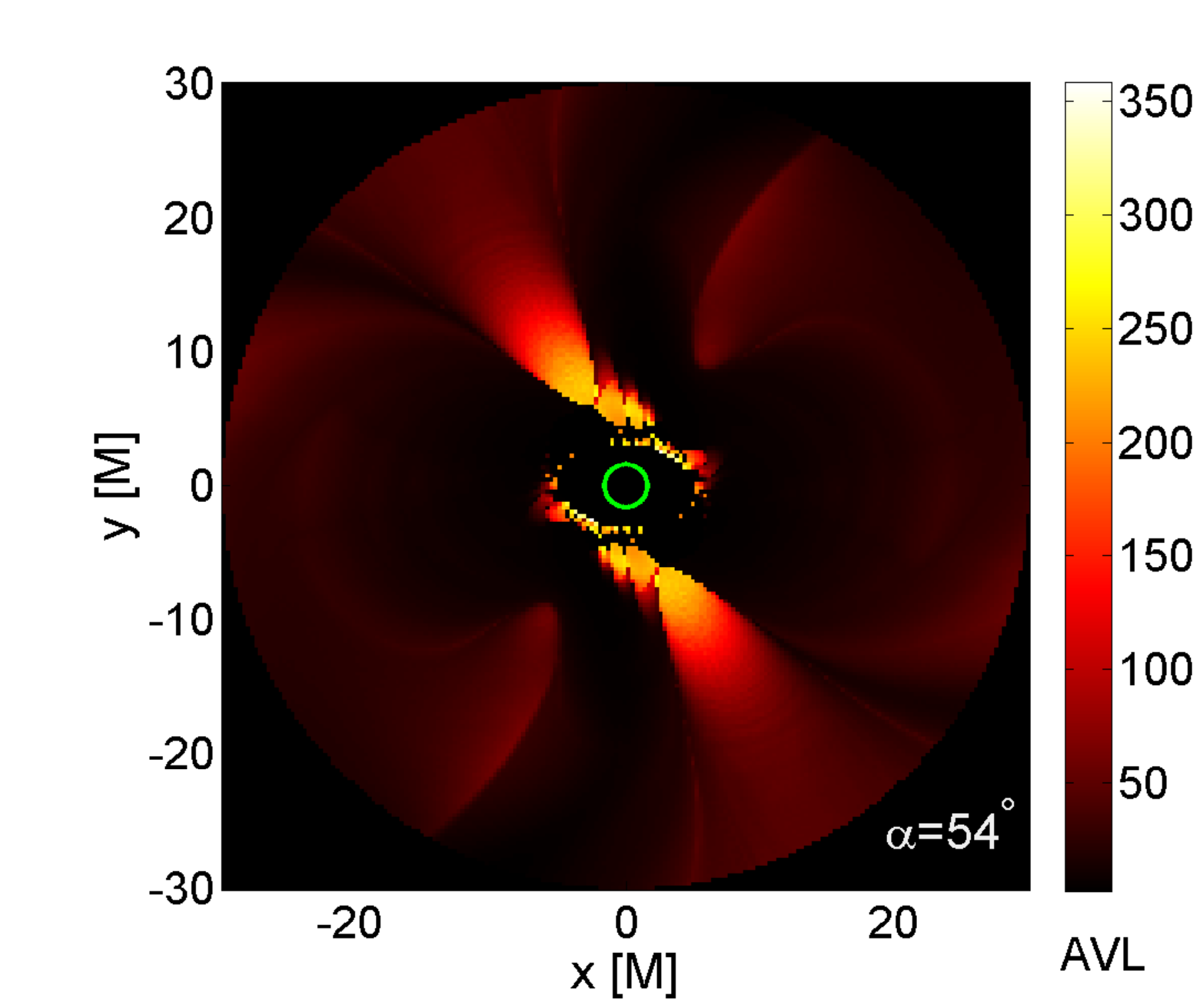}
\includegraphics[scale=.34, trim={12mm 0 2mm 0},clip]{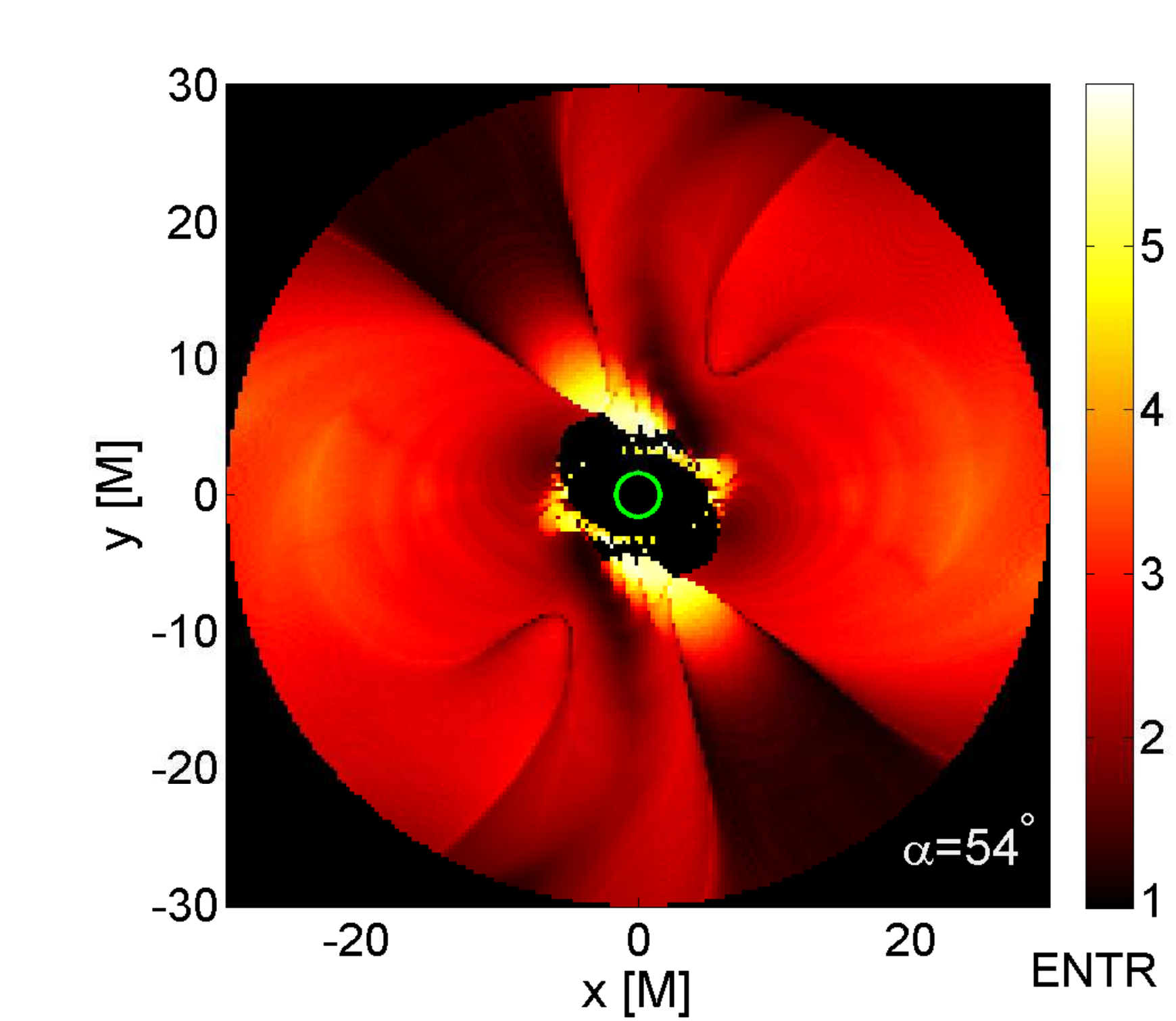}
\includegraphics[scale=.34, trim={13mm 0 2mm 0},clip]{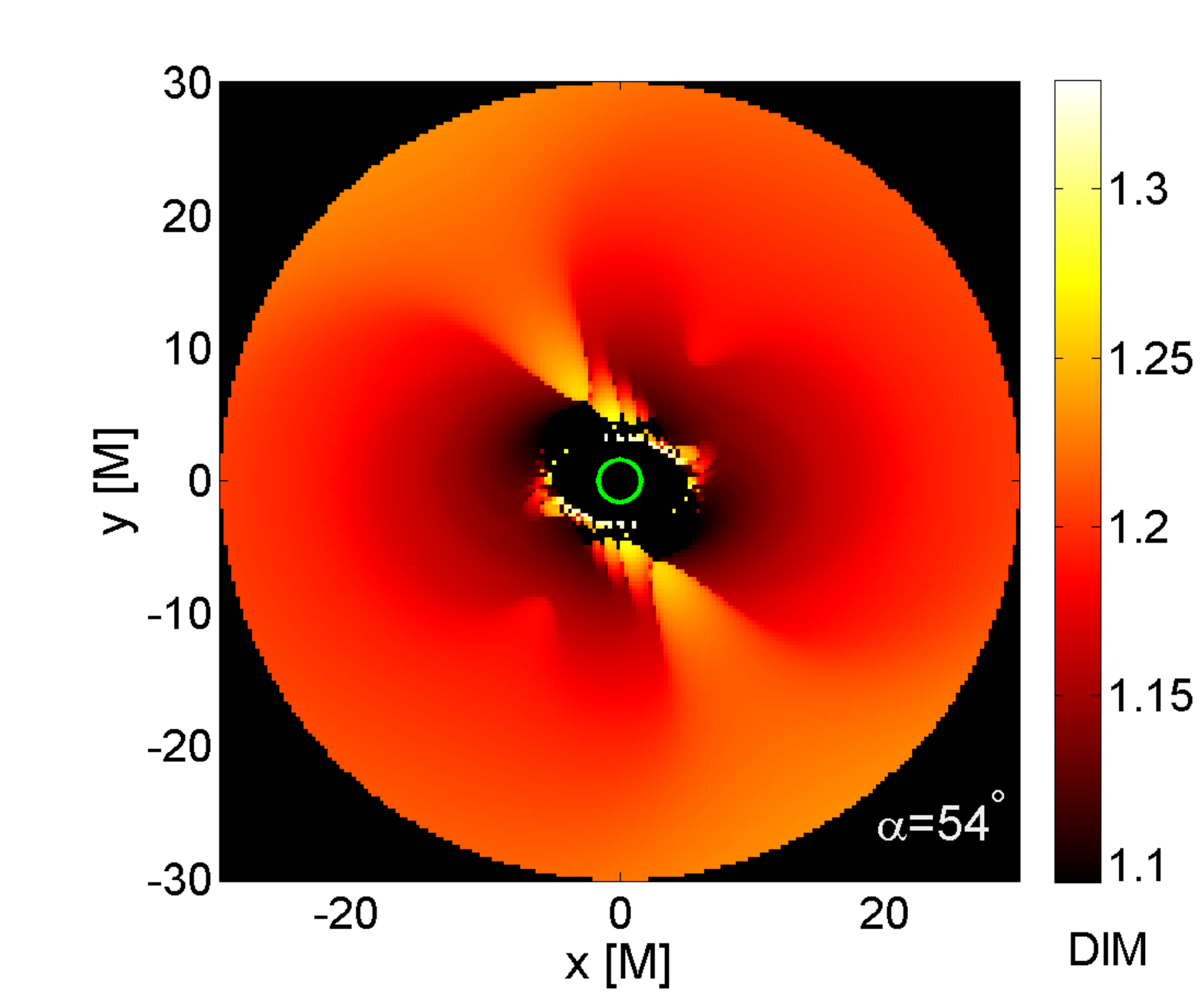}
\caption{Several chaotic indicators evaluated for the trajectories in the escape zones with $qB=-5$, $a=0.98$ (same choice as in Figure~\ref{escape_acc}). Shown are the average length ($\mathrm{AVL}$) of the diagonal lines in the RP, the Shannon entropy ($\mathrm{ENTR}$) of the probability distribution of the diagonal line lengths in the RP, and the box-counting dimension $\mathrm{DIM}$.}
\label{escape_indicators}
\end{figure*}

\begin{equation}
\label{rpdef}
\mathbf{R}_{ij}(\varepsilon)=\Theta(\varepsilon-||\vec{x}(i)-\vec{x}(j)||)\;\;\;
i,j=1,...,N,
\end{equation}
where $\varepsilon$ is a predefined threshold parameter, $\Theta$ is the Heaviside step function, and the vector time series $\vec{x}(t)$ of length $N$ represents the analyzed segment of the trajectory in the phase space. The standard Euclidean norm $L^2$ is applied to evaluate distances after the normalization of individual time series of vector components to zero mean and unit variance, so that they contribute proportionally to the distance.

The recurrence matrix may be directly visualized in recurrence plots (RPs), which are useful for a qualitative inspection of dynamics and intuitive detection of chaos. Previously, RPs were applied in the context of relativistic astrophysics for simulated data \citep[see, e.g.,][]{kopacek10,semerak12,kovar13} and also in X-ray astronomy for the analysis of the light curves of black hole binaries \citep{sukova16a,sukova16b}, while applications in gravitational-wave astronomy have been discussed recently \citep{lukes18,zelenka20}.

Various statistical measures of RPs are employed for recurrence quantification analysis (RQA), which allows a systematic survey of dynamics. A prominent feature of RP to analyze is the presence of diagonal lines. In particular, the distribution of length $l$ of diagonal lines in an RP is given by the histogram $P(\varepsilon,l)$ as
\begin{eqnarray}
\label{diaghist}
\nonumber P(\varepsilon,l)&=&\sum^N_{i,j=1}(1-\mathbf{R}_{i-1,j-1}(\varepsilon))(1-\mathbf{R}_{i+l,j+l}(\varepsilon))\\
 &&\times\prod_{k=0}^{l-1}\mathbf{R}_{i+k,j+k}(\varepsilon).
\end{eqnarray}

The average length ($\mathrm{AVL}$) of the diagonal lines is evaluated as
\begin{equation}
\label{Lav}
\mathrm{AVL}\equiv\frac{\sum^{l_{\rm{max}}}_{l=l_{\rm{min}}}lP(\varepsilon,l)}{\sum^{_{\rm{max}}}_{l=l_{\rm{min}}}P(\varepsilon,l)},
\end{equation}
where only diagonal lines of length at least $l_{\rm{min}}$ count, and $l_{\rm{max}}$ is the length of the longest line (except the line of identity on the main diagonal).

The recurrence indicator $\mathrm{ENTR}$ is defined as the Shannon
entropy of the probability $p(\varepsilon,l)=P(\varepsilon,l)/N_{l}$ of
finding a diagonal line of length $l$ in an RP,
\begin{equation}
\label{entr}
\mathrm{ENTR}\equiv-\sum_{l=l_{\rm{min}}}^{l_{\rm{max}}}p(\varepsilon,l)\ln{p(\varepsilon,l)},
\end{equation}
where $N_{l}$ is the total number of diagonal lines:
$N_{l}(\varepsilon)=\sum_{l\geq{}l_{\rm{min}}}P(\varepsilon,l)$.

Other RQA measures based on the presence of diagonal and vertical lines in an RP may be define; however, here we only employ the indicators $\mathrm{AVL}$ and $\mathrm{ENTR}$. Definitions and properties of other RQA indicators may be found in a review by \citet{marwan07}.

Besides the recurrence analysis, we employ a standard method for characterizing fractal sets in $n$-dimensional metric space, and we evaluate the Minkowski–-Bouligand (box-counting) dimension $\mathrm{DIM}$, defined as 

\begin{equation}
\label{boxc}
\mathrm{DIM}=\lim_{\delta\rightarrow 0} \frac{\log P(\delta)}{\log(1/\delta)},
\end{equation}
where $P(\delta)$ is the number of $n$-dimensional boxes with side $\delta$ required to cover the set. For a complete definition of the box-counting dimension, its properties, and its relation to other fractal dimensions, see, e.g., \citet{falconer03}. The box-counting dimension of the fractal escape boundaries shown in Figures~\ref{escape_detail1} and \ref{escape_detail2} could be evaluated; however, we calculate the indicator of the trajectories directly in order to compare their values within the escape zones. A similar application of the box-counting dimension and other chaotic indicators was recently presented by \citet{panis19}.   

We compute the chaotic indicators $\mathrm{AVL}$, $\mathrm{ENTR}$, and $\mathrm{DIM}$ for the same set of orbits as presented in Figure~\ref{escape_acc} (i.e., for parameters $qB=-5$ and $a=0.98$ with inclinations $\alpha=45\degree$ and $\alpha=54\degree$, respectively); however, we use a shorter segment of the trajectory ($\lambda_{\rm fin}=300$ instead of 1000), and the trajectories are now integrated with the fixed time step as required to perform recurrence analysis correctly. Only a radial coordinate of the trajectory is employed for the analysis; a single time series $r(\lambda)$ with 1000 data points is used as a input. The box-counting dimension is computed with the \texttt{BoxCountfracDim} function,\footnote{Tan H. Nguyen: Hausdorff (Box-Counting) Fractal Dimension with multi-resolution calculation, available at \href{https://www.mathworks.com/matlabcentral/fileexchange/58148-hausdorff-box-counting-fractal-dimension-with-multi-resolution-calculation}{MATLAB Central File Exchange (File ID: \#58148)}.} and RQA indicators are evaluated with the \texttt{CRP Toolbox} \citep{marwan07}. The threshold parameter is set to the fixed value $\varepsilon=0.2$, the minimal length of the diagonal line $l_{\rm{min}}=2$, and embedding is not used. All calculations are performed in \texttt{Matlab R2014a}. 

In Figure~\ref{escape_indicators}, we present color-coded values of indicators of trajectories within the escape zones. Regarding the case of inclination $\alpha=45\degree$ (upper panels of Figure~\ref{escape_indicators}), we observe that the escape zone is clearly distinguished by the values of the indicators. Moreover, the values of the box-counting dimension $\mathrm{DIM}$ (top right panel) to some degree correspond with the values of the final Lorentz factor (left panel of Figure~\ref{escape_acc}). Highly accelerated trajectories tend to have higher $\mathrm{DIM}$, and vice versa. The values of the RQA indicators $\mathrm{AVL}$ and $\mathrm{ENTR}$ do not entirely follow this trend; however, they still clearly distinguish the escaping orbits within the escape zone (especially in the case of $\mathrm{AVL}$). The case of $\alpha=54\degree$, where all three classes of escape zone are present, is analyzed in the bottom panels of Figure~\ref{escape_indicators}. Comparison with the right panel of Figure~\ref{escape_acc} shows that while the shape of the primary escape zone is clearly resolved, only the inner parts of the secondary escape zone are recognized by the increased values of the indicators, and the shape of the tertiary escape zone cannot be distinguished.

The behavior of the chaotic indicators of the trajectories in the escape zones thus corresponds with the observations of the detailed structure of the zones made in Figures~\ref{escape_detail1} and \ref{escape_detail2}. In particular, the comparison with Figure~\ref{escape_indicators} confirms the conjecture that transient chaos plays a dominant role in the dynamics of the primary escape zone. The fractal geometry of this zone directly corresponds with the increased values of chaotic indicators. In the case of the secondary escape zone, only a narrow fuzzy layer of intermixed trajectories was observed on its edge (panels (c) and (d) of Figure~\ref{escape_detail2}), corresponding to increased values of the chaotic indicators in the inner part of the zone. On the other hand, regular dynamics in the tertiary escape zone with well-defined boundaries and no intermixing of trajectories is also confirmed by chaotic indicators, as their values do not increase within this zone.

\section{Conclusions}
\label{conclusions}
We have numerically studied the outflow of matter from the inner region of a magnetized accretion disk triggered by charging of initially neutral accreted particles. As we previously showed in Paper~I, it is possible to model the outflow using the simple setup with the single-particle approach neglecting magnetohydrodynamic effects. Key ingredients that appear sufficient to launch the outflow are the rotation of the central black hole (given by spin parameter $a$) and the large-scale magnetic field. A basic accretion scenario is assumed, and neutral particles are supposed to move initially along (almost) circular Keplerian orbits (turning into freefalling geodesic below the ISCO) and only slowly descend to lower radii until the ionization radius $r_0$ (above or below the ISCO) is reached. Then the particle obtains a nonzero specific charge $q$ and it starts to be affected by the magnetic field. As a result, some plunging particles are stabilized, while some stable orbits turn into plunging. However, the near-horizon escape zone may also develop where the particles escape the attraction of the center and are accelerated along the symmetry axis. Nevertheless, the assumption of perfect axisymmetry (i.e., magnetic field aligned with the spin axis) employed in Paper~I appears to considerably limit the effectivity of the acceleration process. In particular, the maximal final value of Lorentz factor $\gamma_{\rm max}$ saturates at $\gamma_{\rm max}\approx6$, which is attained with $|qB|\gtrapprox 100$ for $a\lessapprox 0.1$. Maximally spinning black holes accelerate the escaping matter only up to $\gamma_{\rm max}\approx2.5$ for $qB\approx-4.5$. However, realistic spin estimates of stellar-mass black holes, as well as supermassive black holes, obtained by various methods from measurements generally lead to moderate-to-high spin values \citep{miller15, daly19,nemmen19, reynolds19}, while high values of the Lorentz factor $\gamma > 10$ are being observed in astrophysical jets of many objects \citep{hovatta09}. 

In the present paper, we have shown that this unrealistic limitation of our model may be removed by relaxing the assumption of axisymmetry. In particular, we have considered the (asymptotically) uniform magnetic field with arbitrary inclination $\alpha$ with respect to the spin axis. It appears that breaking the symmetry by even a small inclination angle ($\alpha\approx1\degree$) is also sufficient to trigger the outflow in cases of high spin and strong magnetization $|qB|$, where the escape is not allowed with the aligned field. As the most important result, we found that considerably higher Lorentz factors may be achieved with the oblique field, including ultrarelativistic velocities with $\gamma\gg1$.

We have employed the method of effective potential (formulated in the frame of the static observer) to determine the necessary conditions for the escape of particles. It appears that for the parallel orientation of the spin axis $z$ and magnetic field component $B_z$, only negatively charged particles may escape (while positively charged  particles escape for the antiparallel orientation). Nonzero spin $a$ and $B_z$ are required for the escape, and the final Lorentz factor is found to be an increasing function of the spin, specific charge $|q|$, and $|B_z|$, while it decreases with the ionization radius $r_0$. However, as these conditions are necessary but not sufficient, we investigate the system numerically to determine the actual location and shape of the escape zones. The emergence and evolution of escape zones  is discussed with respect to the inclination angle $\alpha$. Small-to-moderate inclinations lead to the formation of the primary escape zone which increases and gradually deforms as the inclination rises, breaking the axial symmetry. For high inclinations ($B_x>B_z$), secondary and tertiary escape zones also emerge and grow in size until a certain threshold inclination. As the inclination further increases, the escape zones start to diminish and disappear completely before the perpendicular configuration ($\alpha=\pi/2$) is reached. We have plotted the details of the escape zones in the $r_0\times a$ plane (escape-boundary plots), revealing the complex fractal structure of the primary zone, the narrow fuzzy layer of intermixed trajectories on the edge of the secondary zone, and the well-defined escape boundary of the tertiary zone. The dynamics within the zones was assessed with the set of chaotic indicators (box-counting dimension and two recurrence measures) providing strong evidence of (transient) chaos within the primary escape zone, showing hallmarks of chaotic dynamics in the inner region of the secondary zone and no traces of chaos in the tertiary zone.

We have computed the final Lorentz factor $\gamma$ of escaping particles, confirming that the highest $\gamma$ is achieved in the innermost region of the primary escape zone (with the lowest allowed $r_0$). For a particular (realistically small; $\alpha\approx6\degree$) value of inclination and fixed value of spin ($a=0.98$), we searched for the maximal $\gamma$. Increasing the value of magnetization up to $|qB|=10^3$, we confirmed that (unlike axisymmetric configuration) ultrarelativistic velocities with $\gamma\gg1$ may be achieved. While the acceleration of the particle is actually powered by the parallel component $B_z$, the perpendicular component $B_x$ acts as an extra perturbation that considerably increases the probability of sending the particles on escaping trajectories and also allows the outflow in cases that are excluded in the aligned setup.

Until now, the discussion was held in dimensionless units with all quantities scaled by the rest mass of the black hole $M$. Fixing the value of $M$, we may convert back to SI units; e.g., the magnitude of the magnetic field may be expressed as follows:
 \begin{equation}
 \label{magfield}
B_{\rm SI}=\frac{qB\,c_{\rm SI}}{q_{\rm SI}\left(\frac{M}{M_{\odot}}\right)1472\,\rm{m}},
\end{equation}
where the quantities without subscript ${\rm SI}$ are dimensionless, and $M_{\odot}=1472\,\rm{m}$ is the solar mass in geometrized units. 

If we consider only small inclination angles ($B\approx|B_z|$) and rapidly spinning black holes ($a\approx1$), as we did in Section~\ref{acceleration}, we may link the magnitude of the field with the maximal Lorentz factor achieved by the outflow of particles with a given specific charge $q$ (see Figure~\ref{escape_gamma}). In particular, for the acceleration to relativistic velocities ($\gamma\gtrapprox 2$), we need at least $|qB|\approx 4$. In order to achieve an ultrarelativistic velocity with $\gamma\approx 22$ (which is when the particle's rest energy becomes about $1\permil$ of the total energy), the magnetization of $|qB|\approx40$ is required.

Setting the specific charge of the electron, i.e. $q_{\rm SI}=-1.76\times10^{11}\, \rm{C}\,\rm{kg}^{-1}$, we find that for the stellar-mass black hole of $M=10\,M_{\odot}$, the corresponding magnetic field required for the acceleration to the relativistic velocity ($\gamma=2,\;|qB|=4$) reads $B_{\rm SI}=4.63\times10^{-7}\,\rm{T}$. Large-scale magnetic fields observed in nonthermal filaments in the Galactic Center are supposed to reach the same order of magnitude $B\approx10^{-7}\,\rm{T}$ \citep{ferr10}, although more recent estimates are somewhat lower \citep{yusef13}. Nevertheless, stronger fields sufficient for the acceleration to ultrarelativistic velocities (setting $|qB|=40$ in Equation~(\ref{magfield}) gives $B_{\rm SI}=4.63\times10^{-6}\,\rm{T}$) might be encountered inside molecular clouds observed within the Milky Way \citep{han17}.

The scenario of the outflow of ionized heavy particles and dust grains (carrying considerably lower specific charges than the electron) is also compatible with the conditions encountered in black hole systems (both stellar-mass and supermassive), as previously demonstrated for the particular sources with known masses and magnetic field estimates (Discussion in Paper~I). The presented analysis is not limited to a particular astrophysical object; however, the model keeps its basic astrophysical significance, as the relevant values of the parameters are generally compatible with observations.

Although we considered a simplistic toy model that does not attempt to provide a complete description of black hole jet physics, the analysis provides valuable insight into the role of the ordered magnetic field in the formation and acceleration of the jet. In particular, it shows that even in the particle approximation that neglects magnetohydrodynamic effects, the outflow may be formed and attain ultrarelativistic velocities if these essential ingredients are provided: rotating black hole, large-scale magnetic field, and perturbation of axisymmetry.

\acknowledgements
The authors acknowledge the support from the Inter-Excellence program of the Czech Ministry of Education, Youth and Sports (projects 8JCH 1080 and LTC 18058). VK is thankful for the Czech Science Foundation grant (GA\v{C}R-19-01137J). The Programme for the Development of Scientific Experiments of the European Space Agency is acknowledged for supporting the Czech participation in the eXTP project. Discussions with Martin Kolo\v{s} are highly appreciated.

\appendix
\section{Numerical Integration of Equations of Motion}
\label{appa}
Integration of non-linear Equation (\ref{hameq}) is performed with a multistep Adams--Bashforth--Moulton integrator based on the pre\-dic\-tor--corrector (PECE) method, which is implemented in \texttt{Matlab} as function \texttt{ode113}. An adaptive step size is used, and the local truncation error is controlled by a relative tolerance specified by the parameter \texttt{RelTol}, which we set to the highest allowed precision ($\texttt{RelTol}\approx10^{-14}$). Using this setting, we obtain trajectories with final values of the relative error of energy $E_{rr}$ not exceeding a level of $\approx10^{-5}$. To test whether such precision remains sufficient in the highly nonlinear environment of the oblique black hole magnetosphere, we also employ the Dormand--Prince 8th - 7th order explicit scheme \texttt{ode87}\footnote{Vasiliy Govorukhin: ode87 Integrator, available at \href{https://www.mathworks.com/matlabcentral/fileexchange/3616-ode87-integrator}{MATLAB Central File Exchange (File ID: \#3616)}} \citep{prince81}, which is a high-precision integrator of the embedded Runge--Kutta family with a local error of order $\mathcal{O}(h^9)$. The step size $h$ is adaptive, and with $\texttt{RelTol}=10^{-10}$, we reach a precision of $E_{rr}\approx10^{-7}$ at the end of integration, while the computation time is roughly 100 times longer compared to \texttt{ode113}. The outcome of both integrators is compared on the identical set of trajectories within the particular section of the primary escape zone with the fractal structure. We observe that while the color-coded classification of several individual trajectories changes as we switch the integrators (due to the exponential growth of deviations  in the chaotic domain), the overall structure is conserved, and both schemes deliver a comparable output. On the other hand, if we perform the integration with the lower-order method \texttt{ode45} (Dormand--Prince 5th-4th order scheme; default integrator in \texttt{Matlab} and \texttt{GNU Octave}), the structure changes as a result of global errors of order $E_{rr}\approx10^{-4}$, and large artificial regions of plunging orbits appear in the escape zone. We conclude that while the lower-order method does not perform well enough, the integration scheme \texttt{ode113} provides sufficient precision for the given task. 

We note that we have previously tested the performance of several integrators for the long-term integration of bound orbits in the axisymmetric version of the system \citep{kopacek14b}. Besides the routines described above, we have also employed the implicit $s$-stage Gauss-Legendre symplectic solver \texttt{gls}, which proved superior to \texttt{ode87} in terms of relative error on the long time-scale. For the  applications where very high precision is demanded during the long-term integration of the Hamiltonian system, the method of choice would be a symplectic solver, which is, however, computationally expensive due to implicit formulation and a fixed time step. At present, however, there is no known explicit symplectic solver for the charged particle motion in arbitrary electromagnetic fields. In these cases, the volume-preserving integrator could be used instead \citep{higuera17}. Nevertheless, for the current application (not so long integration of numerous trajectories), the accuracy of the fast multistep integrator \texttt{ode113} remains sufficient.


\begin{thebibliography}{}
\bibitem[Abramowicz \& Fragile(2013)]{abramowicz13} Abramowicz, M.~A., \& Fragile, P.~C.\ 2013, Living Reviews in Relativity, 16, 1
\bibitem[Al Zahrani et al.(2014)]{alzahrani14}Al Zahrani, A.~M. 2014, \prd, 90, 044012
\bibitem[Al Zahrani et al.(2013)]{alzahrani13}Al Zahrani, A.~M., Frolov, V.~P., \& Shoom, A.~A. 2013, \prd, 87, 084043
\bibitem[Babar et al.(2016)]{babar16} Babar, G.~Z., Jamil, M., \& Lim, Y.-K.\ 2016, International Journal of Modern Physics D, 25, 1650024
\bibitem[Bardeen \& Petterson(1975)]{bardeen75} Bardeen, J.~M., \& Petterson, J.~A.\ 1975, \apjl, 195, L65
\bibitem[Bardeen et al.(1972)]{bardeen72}Bardeen, J. M., Press, W. H., \& Teukolsky, S. A. 1972, \apj, 178, 347-370
\bibitem[Bi\v{c}\'{a}k \& Jani\v{s}(1985)]{bicak85}Bi\v c\'ak, J., \& Jani\v{s}, V. 1985, \mnras, 212, 899-915
\bibitem[Blandford et al.(2019)]{blandford19} Blandford, R., Meier, D., \& Readhead, A.\ 2019, \araa, 57, 467
\bibitem[Blandford \& Payne(1982)]{blandford82} Blandford, R.~D., \& Payne, D.~G.\ 1982, \mnras, 199, 883
\bibitem[Blandford \& Znajek(1977)]{blandford77}Blandford, R.~D., \& Znajek, R.~L. 1977, \mnras, 179, 433-456 
\bibitem[Daly(2019)]{daly19} Daly, R.~A.\ 2019, \apj, 886, 37
\bibitem[Falanga et al.(2015)]{falanga15}Falanga, M., Belloni, T., Casella, P., Gilfanov, M., Jonker, P., \& King, A. (editors) 2015, The Physics of Accretion onto Black Holes (Springer-Verlag New York)
\bibitem[Falconer(2003)]{falconer03}Falconer, K. 2003, Fractal Geometry: Mathematical Foundations and Applications (Wiley-Blackwell)
\bibitem[Ferri\`{e}re(2010)]{ferr10}Ferri\`{e}re, K. 2010,  Astron. Nachr., 331, 27-33
\bibitem[Frolov \& Shoom(2010)]{frolov10}Frolov, V.~P., \& Shoom, A.~A. 2010, \prd, 82, 084034
\bibitem[Han(2017)]{han17} Han, J.~L.\ 2017, \araa, 55, 111
\bibitem[Higuera \& Cary(2017)]{higuera17}Higuera, A.~V., \& Cary, J.~R.\ 2017, Physics of Plasmas, 24, 052104
\bibitem[Hovatta et al.(2009)]{hovatta09} Hovatta, T., Valtaoja, E., Tornikoski, M., et al.\ 2009, \aap, 494, 527
\bibitem[Huang et al.(2015)]{huang15}Huang, Q., Chen, J., \& Wang, Y. 2015 Int. J. Mod. Phys. D, 24, 1550054 
\bibitem[Hussain et al.(2014)]{hussain14}Hussain, S., Hussain, I., \& Jamil, M. 2014, Eur. Phys. J. C, 74:3210 
\bibitem[Karas et al.(2014)]{karas14}Karas, V., Kop\'{a}\v{c}ek, O., Kunneriath, D., \& Hamersk\'{y}, J. 2014, Acta Polytechnica, 54, 398-413
\bibitem[Karas et al.(2013)]{karas13} Karas, V., Kop{\'a}{\v{c}}ek, O., \& Kunneriath, D.\ 2013, International Journal of Astronomy and Astrophysics, 3, 18
\bibitem[Karas \& Vokrouhlick{\'y}(1992)]{karas92} Karas, V., \& Vokrouhlick{\'y}, D.\ 1992, General Relativity and Gravitation, 24, 729
\bibitem[Kop\'{a}\v{c}ek \& Karas(2018)]{kopacek18}Kop\'{a}\v{c}ek, O., \& Karas, V. 2018, \apj, 853, 53
\bibitem[Kop{\'a}{\v{c}}ek et al.(2018)]{kopacek18b}Kop{\'a}{\v{c}}ek, O., Tahamtan, T., \& Karas, V.\ 2018b, \prd, 98, 084055
\bibitem[Kop{\'a}{\v{c}}ek \& Karas(2018c)]{kopacek18c}Kop\'{a}\v{c}ek, O., \& Karas, V. 2018c, Fourteenth Marcel Grossmann Meeting - MG14, 1050
\bibitem[Kop\'{a}\v{c}ek \& Karas(2014)]{kopacek14}Kop\'{a}\v{c}ek, O., \& Karas, V. 2014, \apj, 787, 117
\bibitem[Kop{\'a}{\v{c}}ek et al.(2014b)]{kopacek14b} Kop{\'a}{\v{c}}ek, O., Karas, V., Kov{\'a}{\v{r}}, J., et al.\ 2014b, Proceedings of Ragtime 10-13: Workshops on Black Holes and Neutron Stars, 123
\bibitem[Kop\'{a}\v{c}ek et al.(2010)]{kopacek10}Kop\'{a}\v{c}ek, O., Kov\'{a}\v{r}, J., Karas, V., \& Stuchl\'{i}k, Z. 2010, in Proc. Mathematics and Astronomy: A Joint Long Journey, eds. M. de Le\'{o}n, D. M. de Diego \& R. M. Ros, (Springer), pp. 278-287 
\bibitem[Kov\'{a}\v{r} et al.(2013)]{kovar13}Kov\'{a}\v{r}, J., Kop\'{a}\v{c}ek, O., Karas, V., \& Kojima, Y. 2013, Classical Quant. Grav., 30, 025010
\bibitem[Liska et al.(2019)]{liska19} Liska, M., Tchekhovskoy, A., Ingram, A., et al.\ 2019, \mnras, 487, 550
\bibitem[Liska et al.(2020)]{liska19b} Liska, M., Hesp, C., Tchekhovskoy, A., et al.\ 2020, \mnras, doi:10.1093/mnras/staa099
\bibitem[Liska et al.(2018)]{liska18} Liska, M., Hesp, C., Tchekhovskoy, A., et al.\ 2018, \mnras, 474, L81
\bibitem[Lukes-Gerakopoulos \& Kop\'{a}\v{c}ek(2018)]{lukes18}Lukes-Gerakopoulos, G., \& Kop\'{a}\v{c}ek, O. 2018, Int. J. Mod. Phys. D, 27, 1850010
\bibitem[Marwan et al.(2007)]{marwan07}Marwan, N., Carmen Romano, M., Thiel M., \& Kurths J. 2007, Phys. Rep., 438, 237  
\bibitem[McKinney et al.(2013)]{mckinney13}McKinney, J.~C., Tchekhovskoy, A., \& Blandford, R.~D. 2013, Science, 339, 49
\bibitem[Miller \& Miller(2015)]{miller15} Miller, M.~C., \& Miller, J.~M.\ 2015, \physrep, 548, 1
\bibitem[Misner et al.(1973)]{mtw}Misner, C. W., Thorne, K. S., \& Wheeler, J. A. 1973 Gravitation (San Francisco: Freeman)
\bibitem[Nemmen(2019)]{nemmen19} Nemmen, R.\ 2019, \apjl, 880, L26
\bibitem[Novikov \& Thorne(1973)]{novikov73}Novikov, I. D., \& Thorne, K. S. 1973, in Black Holes - Les Astres Occlus, eds. C. De Witt and B. S. De Witt (New York: Gordon \& Breach), pp. 343-450
\bibitem[P\'{a}nis et al.(2019)]{panis19}P\'{a}nis, R., Kološ, M. \& Stuchlík, Z. 2019, Eur. Phys. J. C, 79, 479
\bibitem[Parfrey et al.(2019)]{parfrey19}Parfrey, K., Philippov, A., \& Cerutti, B.\ 2019, \prl, 122, 035101
\bibitem[Penna et al.(2013)]{penna13}Penna, R.~F., Narayan, R., \& Sadowski, A.\ 2013, \mnras, 436, 3741
\bibitem[Penna et al.(2010)]{penna10}Penna, R. F., McKinney, J. C., Narayan, R., Tchekhovskoy, A., Shafee, R., \& McClintock, J. E. 2010, \mnras, 408, 752
\bibitem[Prince \& Dormand(1981)]{prince81}Prince, P.~J., \& Dormand, J.~R. 1981, J. Comp. Appl. Math., 7, 67-75
\bibitem[Reynolds(2019)]{reynolds19} Reynolds, C.~S.\ 2019, Nature Astronomy, 3, 41
\bibitem[Sadowski(2016)]{sadowski16}Sadowski, A. 2016, \mnras, 459, 4397-4407
\bibitem[Semer\'{a}k \& Sukov\'{a}(2012)]{semerak12}Semer\'{a}k, O., \&  Sukov\'{a}, P. 2012, \mnras, 425, 2455
\bibitem[Semer\'{a}k(1993)]{semerak93}Semer\'{a}k, O. 1993, {\em Gen. Relat. Gravit.}, 25, 1041-1077
\bibitem[Shakura(2018)]{shakura18}Shakura, N. (editor) 2018, Accretion Flows in Astrophysics (Springer)
\bibitem[Shiose et al.(2014)]{shiose14}Shiose, R., Kimura, M., \& Chiba, T. 2014, \prd, 90, 124016
\bibitem[Skokos(2010)]{skokos10}Skokos, Ch. 2010, in Dynamics of Small Solar System Bodies and Exoplanets, eds. J. Souchay and R. Dvorak, Lect.\ Notes Phys.\ 790 (Springer: Berlin), pp.~63-135
\bibitem[Stuchl\'{i}k \& Kolo\v{s}(2016)]{stuchlik16}Stuchl\'{i}k, Z., \& Kolo\v{s}, M. 2016, Eur. Phys. J. C, 76, 32
\bibitem[Sukov\'{a} \& Janiuk(2016)]{sukova16a}Sukov{\'a}, P., \& Janiuk, A. 2016, Astron. Astrophys., 591, A77 
\bibitem[Sukov{\'a} et al.(2016)]{sukova16b} Sukov{\'a}, P., Grzedzielski, M., \& Janiuk, A.\ 2016, \aap, 586, A143
\bibitem[Tchekhovskoy(2015)]{tchekhovskoy15}Tchekhovskoy, A. 2015, Astrophys. Space Sc. L. 414, 45-82
\bibitem[T\'{e}l \& Gruiz(2006)]{tel06}T{\'e}l, T. \& Gruiz, M. 2006 Chaotic Dynamics: An Introduction Based on Classical Mechanics, Cambridge University Press 
\bibitem[Tursunov et al.(2018)]{tursunov18} Tursunov, A., Kolo{\v{s}}, M., Stuchl{\'\i}k, Z., et al.\ 2018, \apj, 861, 2
\bibitem[Wald(1974)]{wald74}Wald, R. M. 1974, \prd, 10, 1680
\bibitem[Yusef-Zadeh et al.(2013)]{yusef13} Yusef-Zadeh, F., Hewitt, J.~W., Wardle, M., et al.\ 2013, \apj, 762, 33
\bibitem[Zelenka et al.(2020)]{zelenka20}Zelenka, O., Lukes-Gerakopoulos, G.,  Witzany, V., \& Kop\'{a}\v{c}ek, O. 2020, \prd, 101, 024037
\end{thebibliography}
\end{document}